\documentclass[9pt,twocolumn,twoside]{osajnl}

\journal{stile} 

\setboolean{shortarticle}{false}
\title{Concerns about ground-based astronomical observations: \\ \textsc{Quantifying Satellites' Constellations Damages }}

\author[1]{Stefano Gallozzi}
\author[1]{Diego Paris}
\author[2]{Marco Scardia}
\author[3]{David Dubois}

\affil[1]{stefano.gallozzi@inaf.it, diego.paris@inaf.it, \underline{INAF-Osservatorio Astronomico di Roma (INAF-OARm)}, v. Frascati 33, 00078 Monte Porzio Catone (RM), IT}
\affil[2]{marco.scardia@inaf.it, \underline{INAF-Osservatorio Astronomico di Brera (INAF-OABr)}, Via Brera, 28, 20121 Milano (Mi), IT}
\affil[3]{david.f.dubois@nasa.gov, \underline{National Aeronautics and Space Administration (NASA)}, M/S 245-6 and \underline{Bay Area Environmental Research Institute}, Moffett Field, 94035 CA, USA }
\usepackage{lettrine}
\begin{abstract}
\texttt{Abstract:} This article is a second analysis step from the descriptive arXiv:2001.10952 ([1]) preprint. This work is aimed to raise awareness to the scientific astronomical community about the negative impact of satellites' mega-constellations and estimate the loss of scientific contents expected for ground-based astronomical observations when all 50,000 satellites (and more) will be placed in LEO orbit. The first analysis regards the impact on professional astronomical images in optical windows. Then the study is expanded to other wavelengths and astronomical ground-based facilities (in radio and higher frequencies) to better understand which kind of effects are expected. Authors also try to perform a quantitative economic estimation related to the loss of value for public finances committed to the ground -based astronomical facilities harmed by satellites' constellations. These evaluations are intended for general purposes and can be improved and better estimated; but in this first phase, they could be useful as evidentiary material to quantify the damage in subsequent legal actions against further satellite deployments.  
\end{abstract}

\setboolean{displaycopyright}{true}

\begin{document}

\maketitle

\section{Introduction}
\lettrine[findent=2pt]{\fbox{\textbf{T}}}{ }his work provides evidentiary material to quantify damages and contaminations resulting from the deployment and future service of telecommunications satellite constellations launched into Low Earth Orbit. The aim is to analyse each band suitable for ground-based astronomical observations, taking into account the open electromagnetic windows available to ground-based observations and leaving out from the discussion all bands excluded from observations due to atmospheric opacity at those frequencies, see Fig. 1.

In the second chapter, a general overview on optical ground-based astronomy is given, with technical insight on the Large Binocular Telescope observations; in the third chapter, authors try to analyse the impact of satellites' constellations in terms of Radio Frequency Interference, RFI, for ground-based radio astronomy facilities, concentrating on mitigation techniquess put in place with Iridium Constellation RFI at the Arecibo radio observatory; in the fourth chapter, the Cherenkov Astronomy is investigated as well, with particular attention to the largest Cherenkov experiments and observatories in production and those in preparation. The fifth chapter expands on the larger impact of the rapid growth of mega-constellations, their negative impact on the night sky, and what this means as we move forward in a growing era of space commercial activities; the last chapter gives some quantitative conclusion on the potential impact of satellites' constellations on astronomy and environment as well. Legal concerns are also pointed out.\\
Appendix A describes the basic concepts of satellite constellations and operating modes; appendix B gives a basic optical data reduction primer.  
\begin{figure}[htbp]
\centering
\fbox{\includegraphics[width=0.9\linewidth]{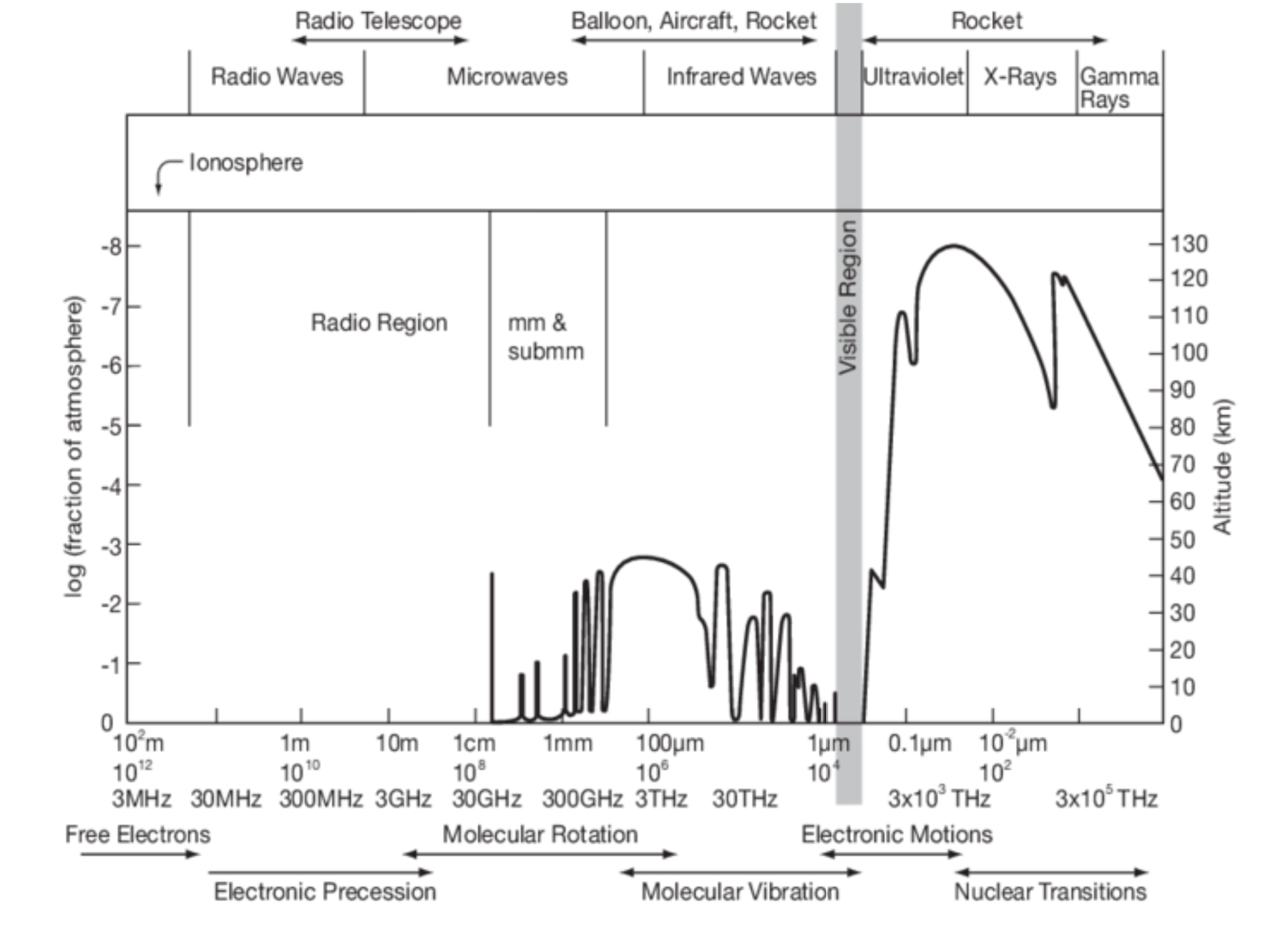}}
\caption{A plot of transmission through the atmosphere versus wavelength, \(\lambda\) in metric units and frequency, \(\nu\), in hertz. The thick curve gives the fraction of the atmosphere (left vertical axis) and the altitude (right axis) needed to reach a transmission of 0.5. }
\label{fig:trasmittanza}
\end{figure}

\section{Damages in optical astronomy}
\label{sec:optdamages}

For professional astronomical observations the number of \textit{satellites above 30 degrees over the horizon} is expected to vary in terms of observing ground-based latitude and can be from \underline{300 to 500 satellites}. Considering Starlink SpaceX's satellites size: 1.1m x 0.7m x 0.7m with 2m x 8m solar panels, the maximum angular size (for the neared orbital shell) is:
\begin{equation}
\theta \propto tan^{-1} \left( \frac{9.1\:m}{alt.} \right)= \begin{bmatrix} 
 0.0015\:deg & 340\:km           \\
 0.0009\:deg & 550\:km           \\
 0.0004\:deg & 1150\:km          \\
\end{bmatrix} 
\label{eq:refname1}
\end{equation}
Where $alt. = (h_{satellite}-h_{observatory})$.\\
Depending on the angular size of Starlink satellites (1) and the pixelscale of the astronomical detectors (their PSF within the optics convolutions and sky seeing of the observing night) Starlink satellites can be approximately considered point-like sources or extended.\\
Depending on the Reference System chosen, absolute Earth center system, ERS and observatory Local system, LRS, see APPENDIX C, the satellites' orbital velocity, \(V\), and the angular velocity, \(\omega\),  can be estimated by:
\begin{equation*}
V=\sqrt{\frac {G*M_E}{R_E+altitude}}\:\:
\label{eq:velocity}
\end{equation*}
\begin{equation}
\begin{matrix}
\omega_{ERS}\propto\frac{1}{R_E+alt.}*\sqrt{\frac{G*M_E}{R_E+alt.}}\\
\omega_{LRS}\propto\frac{R_E/alt.}{R_E+alt.}*\sqrt{\frac{G*M_E}{R_E+alt.}} \\
\end{matrix}
\label{eq:angvel}
\end{equation}
From (2) it is possible to compute the Starlink satellites' orbital velocity depending on the orbital shell, see Table 1.\footnote{There is no information about the orbital velocity of OneWeb satellites, but it is possible to extend the discussion considering the whole OneWeb fleet as the Starlink satellites at 1,150Km altitude.} 
\begin{table}[htbp]
\centering
\caption{\bf Starlink orbital parameters in ERS and LRS}
\begin{tabular}{cccc}
\hline
alt.\:[km] & $V\:[Km/s]$ & $\omega_{ERS}\:[deg/s]$ & $\omega_{LRS}\:[deg/s]$ \\
\hline
$340$ & $7.7$ & $0.066$ &1.2\\
$550$ & $7.6$ & $0.063$ & 0.7\\
$1150$ & $7.3$ & $0.055$ & 0.3\\
\hline
\end{tabular}
  \label{tab:radialvelocity}
\end{table}

For high exposure astronomical images, the radial velocity is big enough to leave trails on detectors. Depending on the time that a satellite remains on a pixel the trail can be saturated or not. This is a tricky situation, but the greater the radial velocity (the lower the orbital shell), the less the time that satellites remain on pixels, saturating them.\footnote{Luminosity is a function of \(\omega\): \\ \text{          - for tracking object: }  \(L(\omega)\propto  \omega^{-2} \)\\ \text{          - for streaked/trailed objects: } \(L(\omega)\propto  \omega^{-1.5}\) \\ This is a general law to compute the \underline{effective satellite magnitudes} in detectors.} It is not necessary to compute the effective magnitude of trails because the scientific content can be preserved only with a masking procedure performed directly on images; in this context the effective trails magnitude is superfluous information.   \\
\begin{figure}[htbp]
\centering
\fbox{\includegraphics[width=0.9\linewidth]{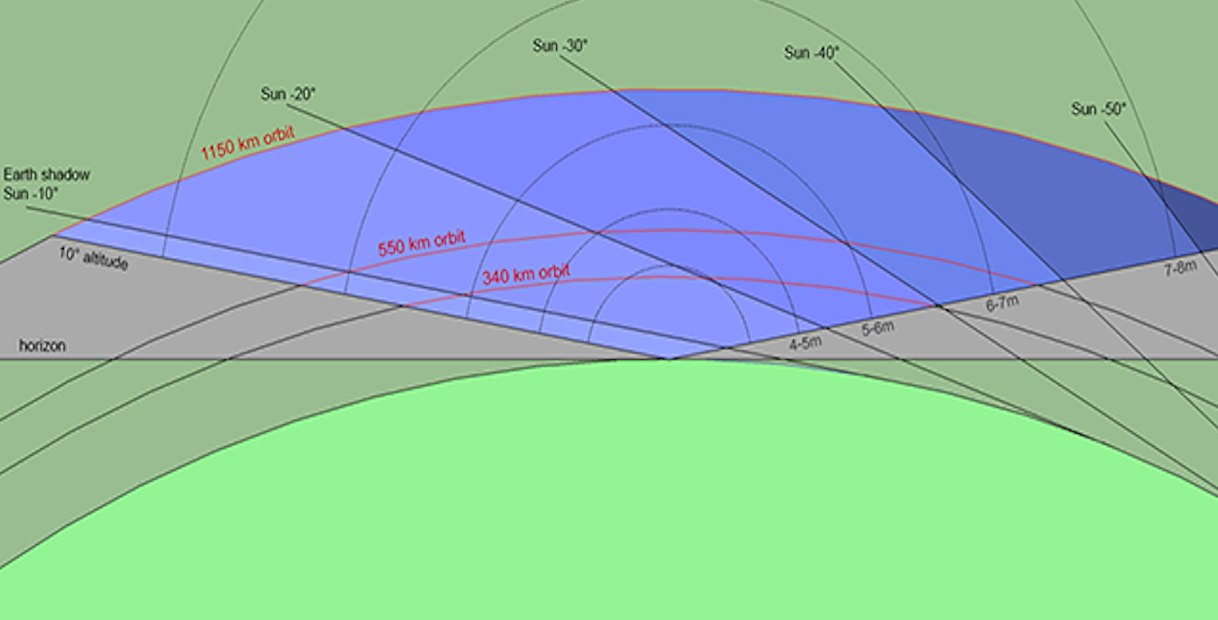}}
\caption{Illumination factor depending on the Sun altitude of the three orbital shells for Starlink satellites. For OneWeb satellites the expected illumination fraction will be the same as the highest Starlink orbital shell, see [28].}
\label{fig:illumfrac}
\end{figure}

Fig. 2 illustrates how Starlink orbital shells (shown in red) are illuminated by the Sun when it is at different altitudes below the horizon. The brightness of satellites as seen from the Earth, depends on their geometric visibility, their orbital inclination, altitude and season. In particular, satellite brightness depends on the angle between the Sun, the satellite and the observer, together with the albedo properties of that satellite. In particular circumstances satellites can flare because of the sunlight glinting off flat surfaces of satellite bodies, creating flashes of light.

Satellites have complex bodies, and one is different from another; what is common to each satellite is the need to be powered by solar panels, so that those reflective flat panels act as mirrors causing specular reflections, that, when pointed toward Earth, cause extremely bright flashes, called \underline{flares}. Each satellite flare can reach to nearly the 5th magnitude, and can be considered a random feature, occurring at random times, when the satellite is illuminated. This damaging behavior is proportional to the number of satellites in the sky: \textit{more satellites means more flares seen on Earth}.\\ 

\begin{figure*}[t]
\centering
\fbox{\includegraphics[width=0.9\linewidth]{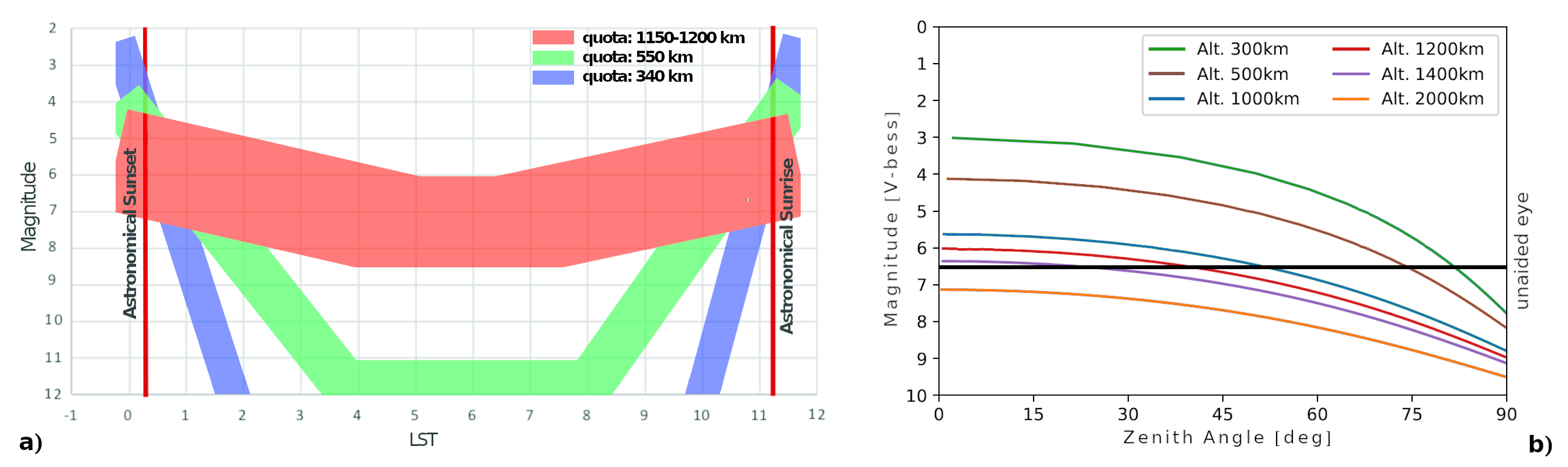}}
\caption{a) Apparent magnitude of satellites during an observing night depending on the altitude. b) Apparent magnitude as a function of zenith angle for different altitudes.}
\label{fig:maglim}
\end{figure*}
The Iridium constellation can be used as a representative constellation to quantify a mean number of flares in observations: Consider 75 satellites (66 that are operative) at 800Km altitude. Each flare illuminates \(\sim 10Km^2\) of area on the ground. The frequency of Iridium flares at 5th magnitude is about four times per week, so \underline{\(\sim\, 0.008\)} flares per day per satellite. The mean number of expected flares for each constellation will be:  \underline{\(\sim\,483\)} ( + facebook flares: , \(\sim\,300\) per 40k sats, \(\sim\,750\) per 100k sats).\\
Instead the number becomes greater if are considered satellites' flares within the 8th magnitude:  \underline{\(\sim\, 0.08\) flares per day per sat}, so that the total mean number of flares are: 50,000 (+ facebook flares TBC!).\\

\begin{tabular}{ c c }
\hline
Constellation & N.flares within 5th mag \\
SpaceX & $\sim\,320$ \\
OneWeb & $\sim\,40$ \\
Telsat & $\sim\,4$ \\
Amazon & $\sim\,25$ \\
Samsung & $\sim\,36$ \\
Kepler & $\sim\,1$ \\
Roscosmos & $\sim\,5$ \\
Chinese Aerospace & $\sim\,2$ \\
Boeing & $\sim\,24$ \\
S\&S Global & $\sim\,1-2$ \\
SatRevolution & $\sim\,8$ \\
CASC & $\sim\,2$ \\
LuckyStar & $\sim\,1$ \\
Commsat & $\sim\,6$ \\
AstroTech & $\sim\,4-5$ \\
\hline
\end{tabular}
\\
When satellites are illuminated by the Sun, depending on the observing night, the observational zenith angle, and the satellite altitude, illumination can reach an apparent magnitude which is capable of saturating astronomical detectors. Also with fainter higher altitude satellites, the mean brightness is enhanced with the cumulative sum of all flares, see Fig. 3a and 3b. These are not only simulations; since the first Starlink was  launched, a direct photometric measurement of 5th Vmag was performed by T. Tyson, as shown in Fig. 10.\\

From the plot it is possible to see that the lower the Sun is, only the more distant satellites will be illuminated.  At certain stages the lowest shells to naked eye won’t be visible at all, but the higher shells will be visible in the northern part of the sky. 
\\
Also the swarm of the satellites near the horizon will be mostly invisible due to their distance and atmospheric effects. It should be remarked that the "worst case" will be experienced during summer in the northern hemisphere, in the northern half of the sky where the satellites will be visible during the entire night; though their brightness will probably be lower than the bright overhead passes after sunset and before sunrise. \\
\begin{figure*}[htbp]
\centering
\fbox{\includegraphics[width=0.9\linewidth]{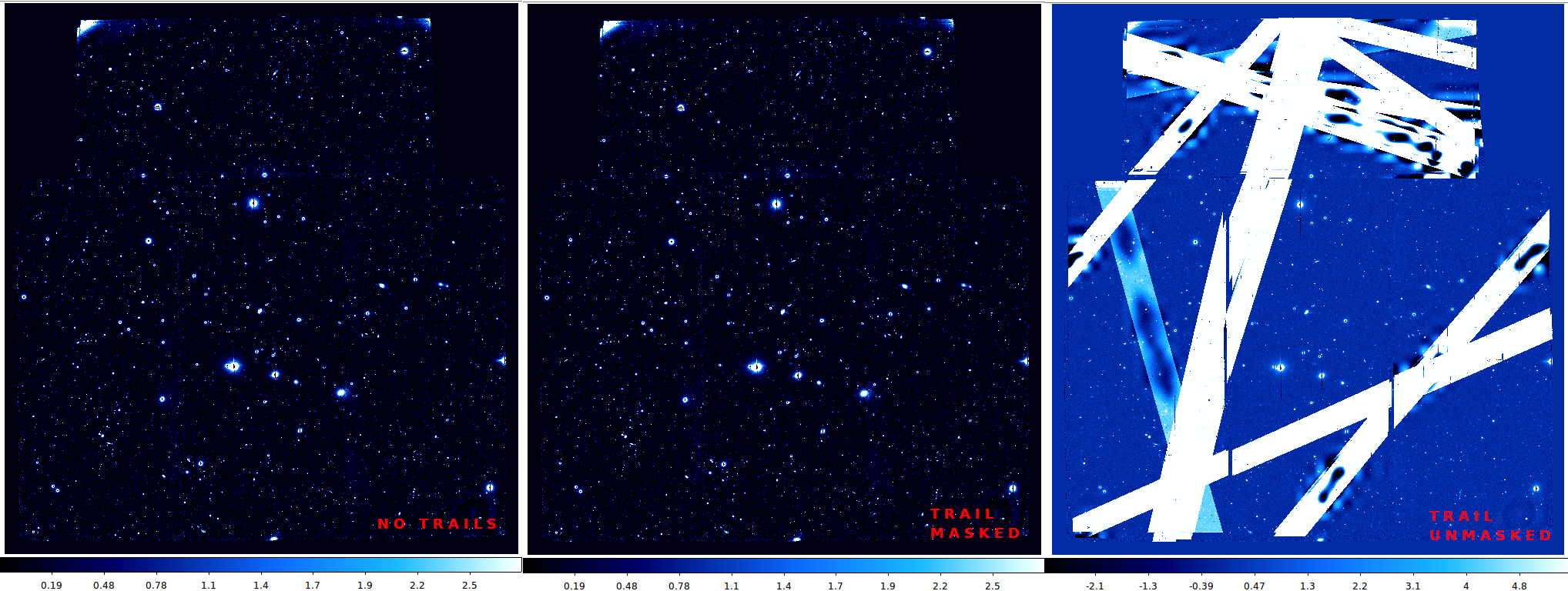}}
\caption{Comparison among different reductions of the same LBC field. The damage to the observations should be evaluated not by the fact that trails can be adjusted or fixed with a little skill and making use of good  reduction techniques, but to the rate of appearance of trails in final images.
}
\label{fig:mosaics}
\end{figure*}

The "best case" will be during the northern hemisphere winter, at midnight when the sky will be virtually free of any satellites, except for the horizon.\\

What is obscured and not visible with the unaided eye \((Maglim_{V-bessel}=6.5)\) is observable with any astronomical imager. In particular, the brightness of the trail depends on the visual limiting magnitude of detectors, and the filter used for the astronomical observation. The night period also interferes with brightness of satellites because astronomical observations near twilight will be crowded with 340Km altitude very bright satellites, while after the astronomical sunset the most light pollution will be originated by the 1,150Km altitude satellites.

Only for large area Field of View (FOV), and very short exposures it can be possible to observe elongated satellites segments instead of trails. Also during the dark night it is estimated that trails leaved by satellites' constellations will be bright enough to saturate modern detectors on large telescopes, see Fig. 3b. \\

As shown later, in Wide-field scientific astronomical observations will be severely affected; in the cases of modern fast wide-field surveys there will be simultaneous multiple trails inside their FOV. \\

Instruments with a smaller field of view would be less affected in terms of number of trailing satellites, but even few trails will damage observations enormously. \\
Depending on the scopes of the work it is possible to use a global reference system, ERS or a local one, LRS. By choosing a good reference system, it is possible to make calculations from the point of view of the local observer, LRS or as the observatory was placed at the center of the Earth, ERS. \\While LRS allows more robust computations it suffers of a completeness problem in the sample of satellites, because orbital configuration of some constellations are not well known so that this approach could led to an underestimate of the total number of satellites. On the contrary LRS allows to easily compute the right number of intersecting satellites at each zenith angle depending on the observatory latitude and altitude. Instead ERS is a generic reference which permits to include all satellites sample in the formulas and allows to make more generic of mean number of satellites in a square degree. So ERS acts like the observatory would be placed at the center of the Earth and so takes care only to evaluate how much satellites intersect the instrument FOV, no matter on the satellite's orbital configuration and the observatory altitude. As seen from table n.1 radial velocity is very different but in contrary the mean density is inversely proportional, so that the mean number of satellites in FOV is similar, see APPENDIX C for details.    

\subsection{The Large Binocular Telescope "test-case"}
To estimate the damage on professional optical astronomical images, authors concentrated on a small set of data coming from the Large Binocular Camera, LBC, see [6], located at the fast first focus of the Large Binocular Telescope, LBT at Mount Graham in Arizona, see [5]. 

A set of public images in V-Bessel Filter (peak around \(525\:nm\)) of the Blue Channel of Large Binocular Camera, LBC has been selected. Over some images of the selected dataset of real LBC observations, several saturated trails were simulated. Brighter trails were simulated over some calibration twilight sky-flats in order to perform a parallel data reduction chain, to investigate the impact of scientific content with altered calibration frames in comparison to reductions performed with unaffected sky-flat images. \\
\begin{figure}[htbp]
\centering
\fbox{\includegraphics[width=0.9\linewidth]{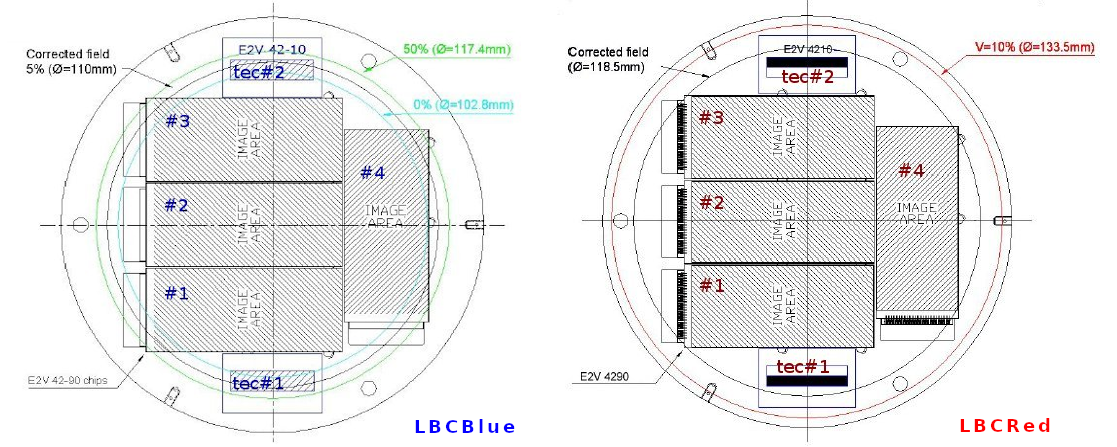}}
\caption{Chips of the Blue and Red channel in the Large Binocular Camera at prime focus of the Large Binocular Telescope, Mt. Graham, AZ.}
\label{fig:lbc}
\end{figure}

The LBC camera has a FOV of \(\sim 25\:arcmin\,\times\,\sim 23\:arcmin\:(\to\,diagonal\,\sim 33\:arcmin )\), see Fig. 5. At the f/1.47 LBC focus, the focal plane scale is \(16.9\, arcsec/mm\), and so, the average pixel scale for both LBC Blue and LBC Red is \(0.2255\, arcsec/pixel\), see Fig. 5.\\ Each LBC FOV is crossed by Starlink satellites in a very short time: \\
\begin{tabular}{ccc}
\hline
\: &\underline{ in ERS:} &\\
36\,ms\:(at\,340\,Km) & $34\,ms\:(at\,550\,Km)$ & $30\,ms\:(at\,1150\,Km)$ \\
\: & \underline{in LRS:} &\\
0.66\,s\:(at\,340\,Km) & $0.38\,s\:(at\,550\,Km)$ & $0.16\,s\:(at\,1150\,Km)$ \\
\hline
\\
\end{tabular}
\\
The LBC projected area is \(25\times23\,arcmin^2=0.16\,deg^2\), so considering the mean density of satellites in sky, \(\rho\) and the solid angle \(\Omega=4\pi * \left(\frac{180^\circ}{\pi}\right)^2=41,253\,deg^2\):
\begin{equation}
\begin{matrix}
\rho_{ERS} \propto\frac{\#sats}{\Omega}\simeq \frac{50,000}{41,253}\simeq 1.2sats\,deg^{-2} \\
\rho_{LRS} \propto\frac{\#sats}{\Omega}\cdot \frac{alt.}{R_E}  \,=\,

\begin{matrix} 
\simeq 0.06\,sats deg^{-2}\:at\, 340Km\\
\simeq 0.1\,sats deg^{-2}\:at\, 550Km\\
\simeq 0.2\,sats deg^{-2}\:at\, 1150Km
\end{matrix}

\end{matrix}
\end{equation}
\\
According to radial velocity in Tab.1, each pixel of LBC camera will be crossed in \(\sim\,1\,ms\); so taking into account the LBC Exposure Time Calculator, see [61], it is possible to compute the saturation magnitude in V-Bessel Filter for the Blue LBC-channel that for \(\sim \, 1\,ms\) is 
\begin{equation*}
M_{saturation} \simeq -2.5*log_{10}\left( \frac{flux_1}{ExpoTime} \right)+Zp^{airm=0}_{1sec} \simeq \, 4,74
\end{equation*}\\
Where \(flux_1\) is the total flux divided by the total exposure time and \(zp^{airm=0}_{1sec}\) is the magnitude zero point for \(airmass=0.0\) at 1 second of exposure time.\\
Because of the brightness of starlink satellites, as seen in Fig. 3b, each Sun-illuminated satellite will leave a saturated trail in LBC chips; so each saturated trail will be masked and expanded as seen in Fig.6.\\
\begin{figure}[htbp]
\centering
\fbox{\includegraphics[width=0.9\linewidth]{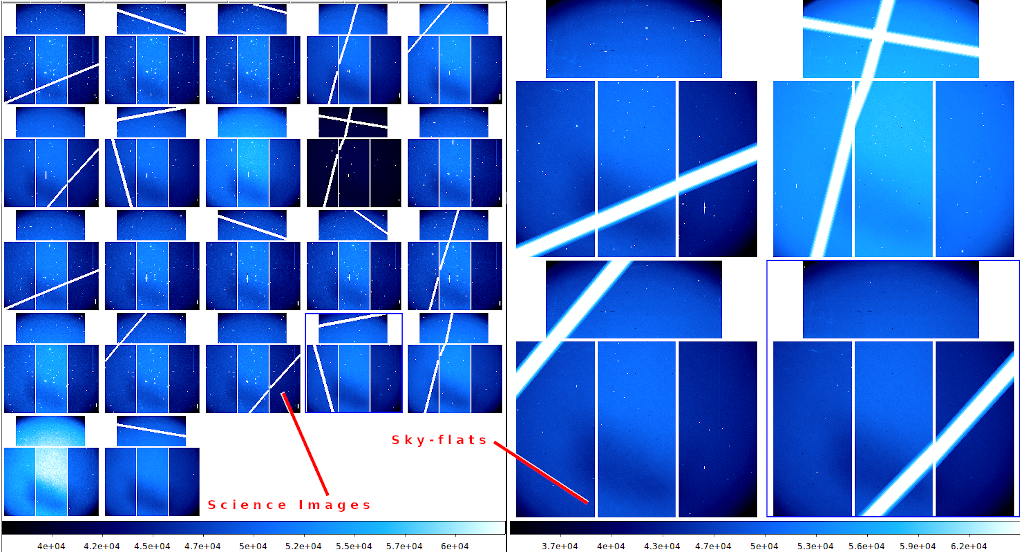}}
\caption{A set of damaged flat fields at twilight on the right and scientific images with trails on the left.}
\label{fig:immaginitrails}
\end{figure}

The number of 50,000 satellites has been taken by a projection of the latest approved project, but the number could be extremely underestimated because each internet provider could, in the future, claim to deploy its own private fleet and send into LEO orbit, see Tab.2.\footnote{Microwave spectrum is divided in different bands:	L-band (1-2GHz),  S-band (2-4GHz), C-band (4-8.2GHz), X-band (8.2-12GHz), K$_u$-band (12-18GHz), K-band (18-26.5GHz), K$_a$-band (26.5-40GHz), Q-band (33-50GHz), U-band (40-60GHz),  E-band (60-90GHz), V-band (40-75GHz), W-band (75-110GHz), D-band (110-170GHz), G-band (110-300GHz), Y-band (325-500GHz).} The aim of this work is to make an estimation on the \underline{worst expectation} even if some projects could be rejected or withdrawn (e.g. Boeing) or not yet confirmed, it is necessary to make a projection with a round number in order to multiply to the related factor as new projects will be included or excluded from the list, that is the reason why in this work authors concentrate on the ERS Reference System.

\begin{table*}[t]
\centering
\caption{\bf Satellite Constellation projects comparison in terms of satellites number, orbital shell altitude, involved bands and foreseen date for service startup (*) if "?" the project has to be confirmed or authorized or even withdrawn.}
\begin{tabular}{|c|c|c|c|c|}
\hline 
\color{blue}\textbf{Constellation Name} & \color{blue}\textbf{num. Satellites} & \color{blue}\textbf{Altitude {[}km{]}} & \color{blue}\textbf{Bands} & \color{blue}\textbf{Service Start}\color{black}\tabularnewline
\hline 
\hline 
SpaceX - Starlink (USA) & 42,000 & 1,150, 550, 340 & K$_u$, K$_a$, V & 2020\tabularnewline
\hline 
OneWeb (UK) & 5,260 & 1,200 & K$_u$ & 2020\tabularnewline
\hline 
Iridium (USA) & 75 & 780 & L & 2018\tabularnewline
\hline 
Telesat (CAN) & 512 & \textasciitilde 1,000 & K$_a$ & 2021-2025\tabularnewline
\hline 
Amazon - Kuiper (USA) & 3,236 & 590, 630, 610 & ? & 2021\tabularnewline
\hline 
Samsung (KOR) & 4,700 & 1,400 & ? & 2021-2030\tabularnewline
\hline 
Lynk (USA) & 36 & 500 & ? & 2021-2023\tabularnewline
\hline 
Kepler Comm. (USA) & 140  & 575 & X, K$_u$ & 2021\tabularnewline
\hline 
\color{red}Facebook Athena (USA) & \color{red} ?40k,100k, 400k? & \color{red}500-550 & \color{red}? & \color{red}2020-2030\color{black}\tabularnewline
\hline 
Roscosmos (RU) & 640 & 870 & L - X & 2022-2026\tabularnewline
\hline 
LeoSat Ent. (USA) & 108 & ? & K$_a$ & ?\tabularnewline
\hline 
C.Aerosp.ScienceTech.Corp. (CHI) & 300+ & \textasciitilde 1,000 & L, K$_a$ & 2022\tabularnewline
\hline 
Boeing (USA) & 3,116 & 1,200 & V, C, K$_a$ & ?\tabularnewline
\hline 
Sky and Space Global (UK) & 200 & ? & L, S & ?\tabularnewline
\hline 
SES (USA) & 42 & ? & K$_a$ & ?\tabularnewline
\hline 
Globalstar (USA) & 48 & 1,400 & S & ?\tabularnewline
\hline 
ViaSat (USA) & 24 & ? & K$_a$, V & ?\tabularnewline
\hline 
Karousel LLC (USA) & 12 & ? & K$_a$ & ?\tabularnewline
\hline 
Sat Revolution (USA) & 1,024 & 350 & R, G B, NIR & 2021-2026\tabularnewline
\hline 
CASC (CHI) & 320 & 1,100 & L, K$_a$ & ?\tabularnewline
\hline 
LuckyStar (CHI) & 156 & 1,000 & S & ?\tabularnewline
\hline 
Commsat (CHI) & 800 & 600 & optical & ?\tabularnewline
\hline 
Xinwei (CHI) & 32 & 600 & C, K$_a$, K$_u$ & ?\tabularnewline
\hline 
Astro Tech (IND) & 600 & 1,400 & C & ?\tabularnewline
\hline 
\color{blue}\underline{TOTAL} & \color{blue}\textbf{63,381 \color{red}(+Facebook?)} & \color{blue}340 <-> 1,400 & \color{blue}ALL & \color{blue}in 10 years\color{black}\tabularnewline\hline 
\end{tabular}
\end{table*}
For the medium size of LBC FOV, the number of satellites producing trails in a mosaic of about one hour of total exposure is \(\simeq\,30\,sats\): \underline{approximately one trail every 120 seconds}; so depending on which kind of observation is planned and which filter is going to be used, for LBC the normality could result in having one to two trails in each scientific exposure RAW-frame.\footnote{This number could be \underline{highly underestimated} since the Facebook CEO, M. Zuckerberg, recently spoke of a service capable of moving ten times the data flux of SpaceX. Thus even considering the same number of SpaceX's satellites, the total amount of telecom (TLC) satellites in LEO could exceed 100,000!}

To test the impact of those trails on LBC professional data, a set of images has been extracted (with a 150 second single exposure in V-BESSEL filter) for a total mosaic exposure of about one hour, see Fig. 6. On those images the same standard reduction was performed in three different situations: 1) \underline{without trails}, 2) \underline{with trails masked} by special algorithms and 3) with \underline{trails unmasked}, see Fig. 4 for data reduction results.\\

It is possible to remark that the final mosaic with masked satellite trails is very similar to the original mosaic without trails in it, but the exposure map is quite different, see Fig. 7. \\

The loss of image depth has repercussions on the magnitude estimations error, their computed value, and finally on their magnitude limit: in the chip areas intersected by trails, the image depth is lower than in areas not intersected, so that different sources near the limiting instrumental magnitude are not detected or just badly computed.  

\begin{figure}[htbp]
\centering
\fbox{\includegraphics[width=0.9\linewidth]{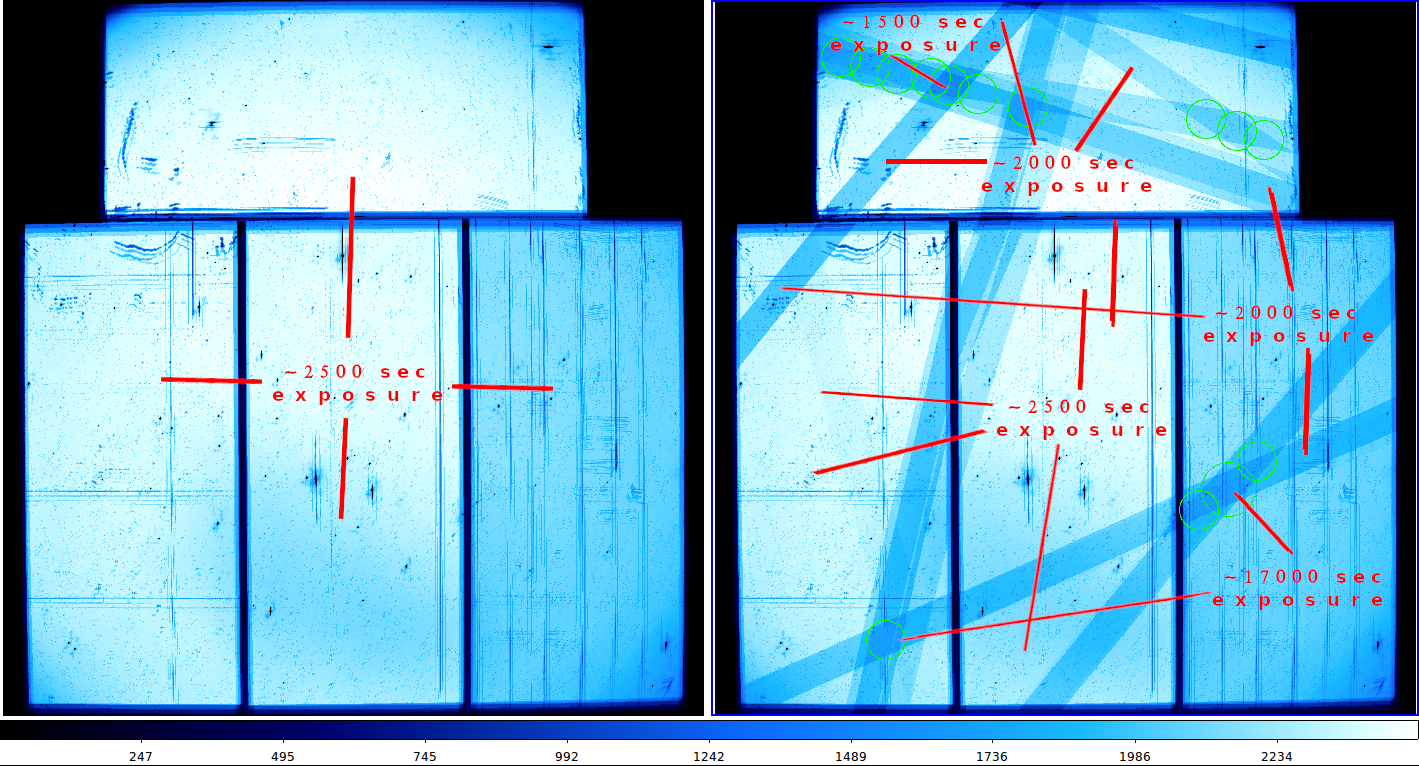}}
\caption{Comparison between clean exposure map and an exposure map of the same field with a lot of trails masked.}
\label{fig:expomaps}
\end{figure}
As described in Fig. 9 the mag computation is seriously affected by the presence of trails, even with the huge effort to mask them, since the precision in magnitude computation decreases, and several sources present a very high deviation from the zero value. This must be considered and more deeply investigated before starting high precision photometry studies. \\
\\
\textit{All the manpower effort and scientific overheads to mitigate the impact of trails on scientific images in the dark-night observations are potentially annihilated by the impact of satellites during the twilight hours.} In particular, each astronomical image, to be usable as science-ready data, needs to be reduced using standard calibration images known as sky-flats. 

\begin{figure*}[t]
\centering
\fbox{\includegraphics[width=0.9\linewidth]{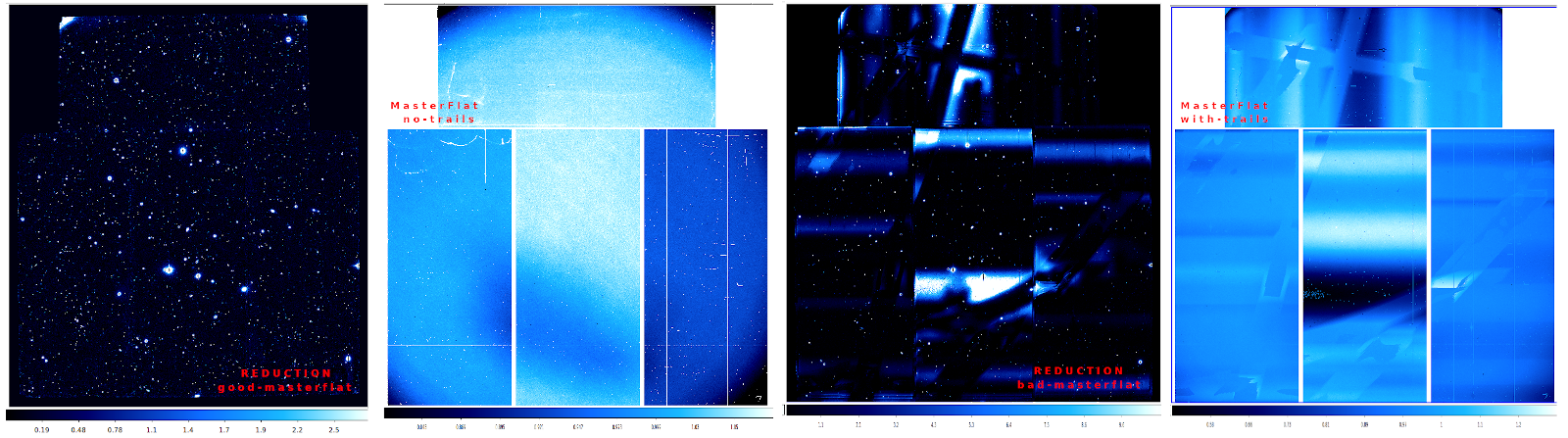}}
\caption{Comparison between a good mosaic generated with pre-processing algorithm using good master flat image and the corresponding reduction using a bad master flat generated with flat frames containing satellite trails.}
\label{fig:reductions}
\end{figure*}
These images are taken with small exposures (about 10 seconds) and are usually combined to obtain a final median stacked pixel by pixel image: this process generates the calibration image known as a \underline{master flat}. \\

Each time a bright LEO satellite (340Km or 550Km altitude) crosses the FOV during a flat exposure, the master flat is contaminated and the flat frame should be excluded by further computations/reductions. \\

If a significant fraction of flat image results are affected by satellite trails, a good sky-master flat cannot be produced. This situation could happen very frequently, because it is known that SpaceX satellites will be equipped with a dedicated navigation laser system that will check and correct the position of each satellite with respect to the other close satellites. This feature is used to maintain the orbital asset in an optimal configuration, and, consequently means that will not be possible to accurately predict the exact position of each satellite, only the approximate location without the intervention of its autonomous guide system.

So, considering that the number of visible satellites is greater during the twilight and that the master flat creation is an automatic standard process, if the prediction of exact satellite position is difficult, the eventuality of interrupted observations during the satellites' passage could be quite impossible, thus the probability to alter flat field images is very high.\\

If a data reduction is performed with a contaminated flat-field, the whole reduction chain is compromised, so the scientific content of the night observation (even with trail masked) becomes damaged or unusable, see Fig. 8 and 9.\\

The flat field operation is crucial in optical astronomical data reduction, since the sensitivities of each pixel in a CCD camera usually varies with time, and it is fundamental to get a good set of flat images before (and after) each observing night in order to better calibrate images, see Appendix B for details.\footnote{The standard reduction pipeline for LBC camera images consists of different steps: 1) chip cross-talk; 2) pre-reduction: debias and flatfielding; 3) background subtraction; 4) astrometric solutions; 5) resampling and coaddition into mosaic final image, see [43] and Appendix B.} 
\\
Some astronomical observatories adopt a different strategy, making use of domes-flat images instead of sky-flats: these images can correct pixel-to-pixel sensitivities, but unfortunately they introduce light illumination additive gradient patterns, which are very difficult to correct/erase in a second processing step since the flat-fielding operation represents a multiplicative factor in data reductions pipelines. \\

It is clear that the final magnitude computation will result in higher magnitude differences between a clean reduction mosaic with respect to those with badly trailed master flats, see last plot in Fig. 9. More importantly, using a bad master flat calibration, about 90\% of the sample sources are lost since they are not matched with the original catalog and/or are above the detection threshold. This is a fundamental argument to highlight the need to obtain good calibration frames to ensure proper data-reduction computation in order to reach the maximum limiting magnitude of the instrument and detect very faint objects.\footnote{Another important limitation of of such a loss of sources is that another data reduction step becomes very difficult: find the astrometric solution for the mosaic.} \\

\textit{This is the most serious concern about damages produced by satellite constellations in optical ground-based professional astronomical observations.}

\subsection{What about other instruments and observing methodologies?}

The identification of LBT as a case study of this work depends on the characteristics of the facility, which gives the observer the opportunity to use different instruments with complementary observing techniques.
Not only imaging can be used, but also spectroscopy and Adaptive Optics facilities, as well as interferometric NIR instruments are available. \\

It is clear that satellite passages in the telescope FOV will necessarily produce damages, but as in the imaging, the possibility to mitigate or correct problems can vary, depending on the data and its expected use.\\

In the \underline{IR range} the number of relevant satellites for astronomy is just the total number of satellites in the sky, because the illumination fraction is not important, and satellites also thermally emit under Sun shadow: operating temperatures can reach \(\sim\,300\,K\) and this will produce an IR flux in M and Q-bands. \\

For instance, observations with \underline{IR/NIR} instruments such as LUCI or MODS are taken with a small exposure in order to minimize the thermal sky background which dominates the signal in ground-based IR observations. This produces faint objects over a bright thermal background, and observational techniques are fundamental to produce scientific content. Flat strategy is important, too. Other IR facilities can implement other acquiring techniques (e.g. "chopping" or "nodding" implemented at a few Hz rate on the secondary mirror) and, in case of trails, very short exposure can be easily rejected from the reduction without affectiing the whole scientific content. \\

Moreover, the satellites' trails will be seen as faint sources over the very bright background, and won't be able to saturate detectors in IR/NIR.\footnote{In \(5-20\mu m\) each satellite will emit a considerable amount of thermal radiation.} This, significantly, makes the satellites' damages in IR/NIR astronomy not very relevant, even if trails need to be masked.

In \underline{spectroscopy investigations}, depending on the band, the key features are also exposure time, filter, the grating and dispersing elements, dimension of the instrument FOV and length of slits involved in spectroscopy analysis. In particular, \underline{Slitless spectra} without masks will be highly affected, but depending on the observing strategy, the effect could be mitigated: if small exposure spectra are taken, then those with trails could be excluded from further processing in spectral pipelines, so the real damages will be limited to overhead and adding to each observation a corresponding number of frames to substitute for all exposures lost with trails. In practice, long exposure slitless spectra could be very hard, if not impossible, see Fig. 11.\\

Instead, using a particular mask, if satellites pass behind the mask in a covered region of the field, spectra could be taken without great damage. It is not clear how this can be planned in dedicated scheduling operations before the data-taking, but it is clear that this corresponds to additional costs and losses of efficiency for the whole observatory infrastructure as well.\\

Similar discussions can be extended to \underline{Interferometry} and \underline{Adaptive-Optics} instruments and techniques, since in AO systems the FOV is very small, the same for Multi Conjugate Adaptive Optics (MCAO).\\

\begin{figure*}[t]
\centering
\fbox{\includegraphics[width=0.9\linewidth]{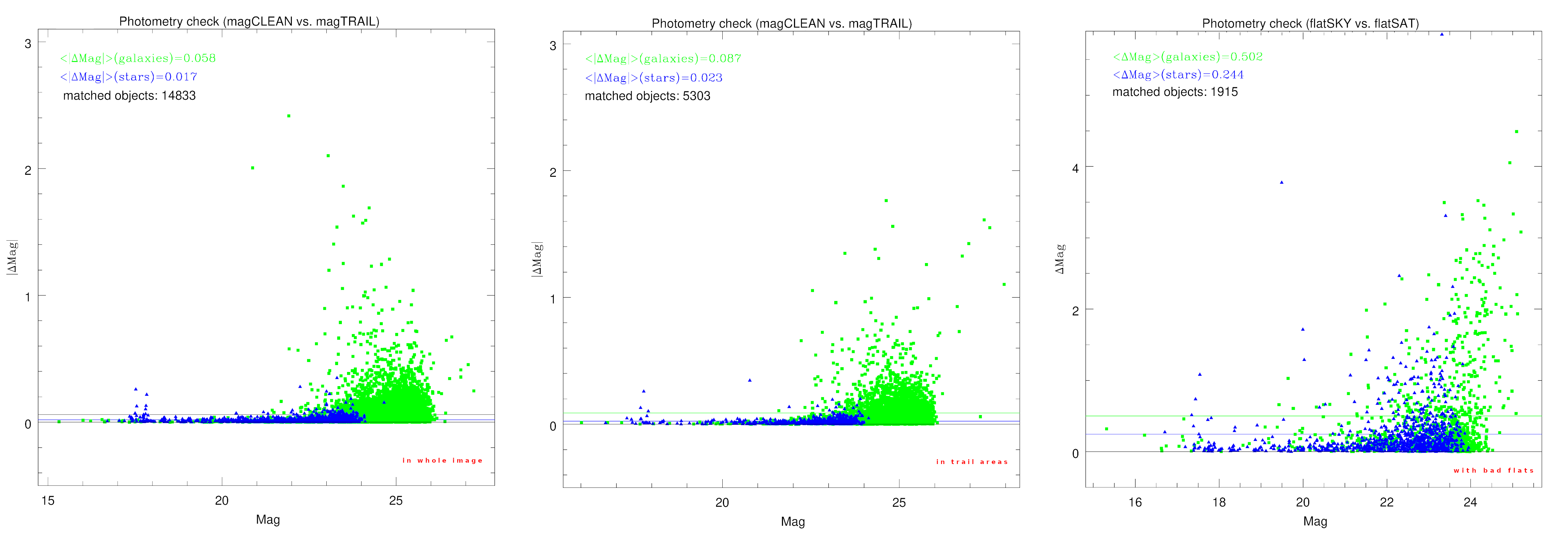}}
\caption{The \underline{left image} plots the absolute \(|\Delta Mag|\) between computed magnitude in V-BESSEL filter from a clean mosaic, and one with trails masked. The \underline{central image} shows the same plot performed only in trail areas. The \underline{right image} plots the absolute magnitude differences between a catalog extracted from a good mosaic generated with a good master flat, and a second catalog extracted from a bad mosaic generated with a bad master flat with trails inside: at least 90\% of sources in the field are not detected.}
\label{fig:plotmags}
\end{figure*}

So in general, in very small Field of View Astronomical Observations, there are no severe damages foreseen by the impact of satellites' constellations.

\subsection{What about Wide Field of View Observatories?}

Regardless of the minor exposure time, single exposure, wide-field survey telescopes will be particularly damaged, because of the large FOV, by the presence of multiple saturated trails within each single camera image:\\

\begin{figure}[t]
\centering
\fbox{\includegraphics[width=0.8\linewidth]{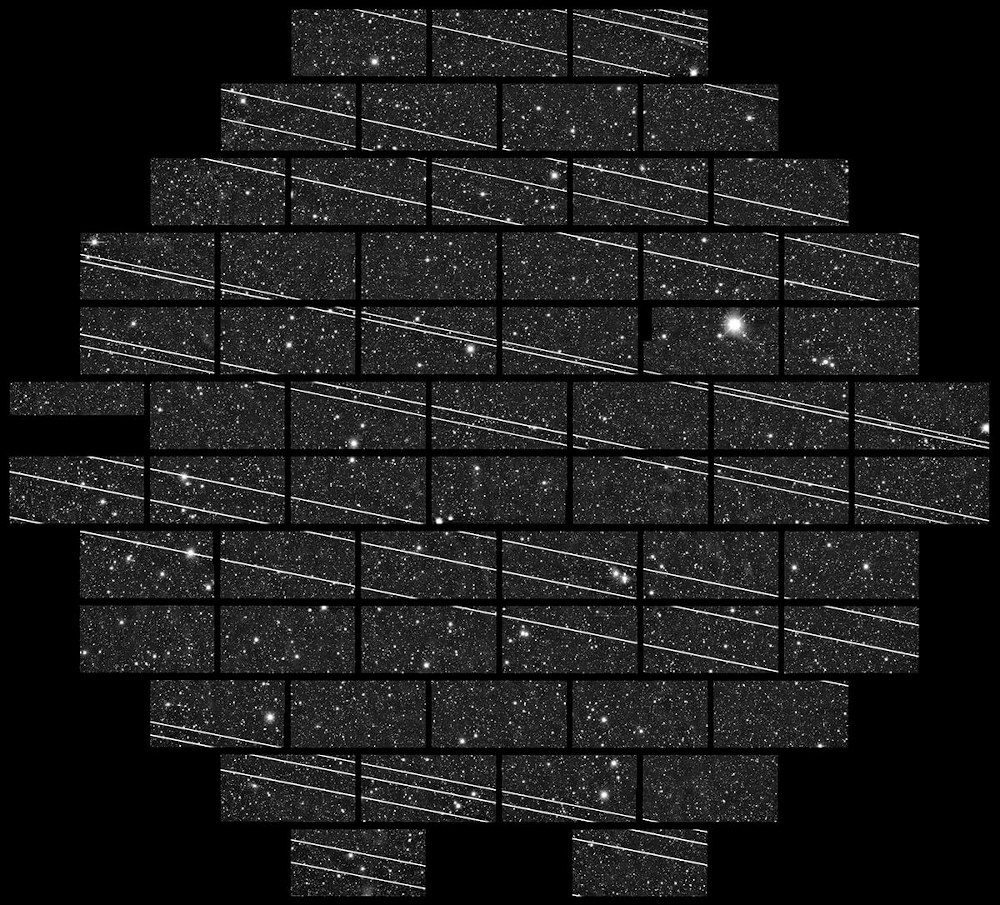}}
\caption{ Starlink satellites visible in a mosaic of an astronomical image (courtesy of NSF’s National Optical-Infrared Astronomy Research Laboratory/NSF/AURA/CTIO/DELVE). The photometric single exposure measurement of this set of trails revealed a 5th Vmag apparent magnitude, from T. Tyson.}
\label{fig:strisciateridotte}
\end{figure}
\begin{itemize}
   \item VST [6], with its 268 MegaPixels camera and a FOV of 1 square degree \(\rightarrow\, 1-2\, simultaneous\, trails\)
   \item LSST [5] (i.e. Rubin Observatory), with a 3.5 degree FOV  \(\rightarrow\, 4\, simultaneous\, trails\)
\item Pan-STARRS [7], with its FOV of 7 square degrees and 1.4 Gigapixel camera \(\rightarrow\,8\,simultaneous\, trails\)
\end{itemize}

\begin{figure*}[t]
\centering
\fbox{\includegraphics[width=0.9\linewidth]{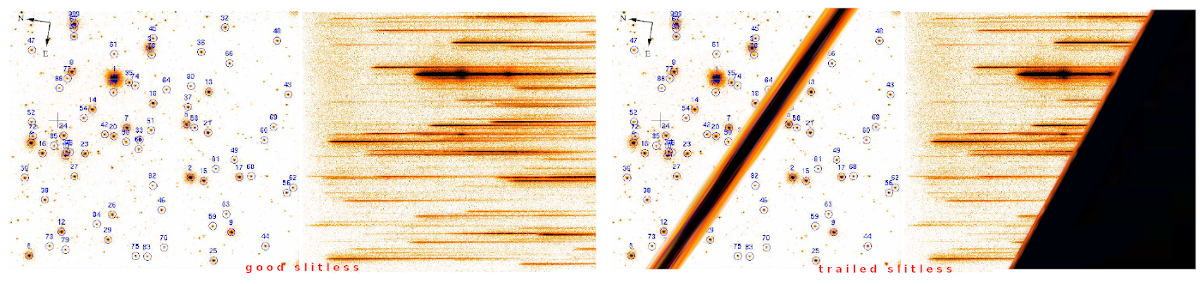}}
\caption{In the two pairs of images, damages from a single trail on a slitless observation are shown.}
\label{fig:slitless}
\end{figure*}

The same LBT problems arising from the flat-fielding procedure are common to large area survey instruments, while for the night sky observations the crucial factor is the exposure time, according to the radial velocity of each satellite in each constellation, see Fig. 10 as a single wide FOV exposure.

\subsection{What about Variability Studies?}

If few satellites cross the sky, variability studies are not really affected, but if the number of such trails increases, occultation probability will also rise. We should talk of occultation both in the case of not-illuminated satellites and in the case of bright trails, because if a particular object is observed, what is computed is the variability of the light function. \\
Considering the radial velocity in the sky of those satellites, the time needed for a point-like source to produce an occultation of the light coming from the astronomical object is very short (fractions of msec). \\

Taking into account any instrument FOV and the probabilistic number of satellites in the sky (and thus in that FOV), the probabilistic chance of hitting an occultation is moderate, \(\sim 1\%\). So all variability studies could be damaged in the same measure of "small" probability up to \(\sim 1\%\), depending on the exposure time of the single observation.\\

In principle, a good occultation timing could even be used to correlate variable astronomical objects as fine tuning of computation methods related to astronomical sources variability detection and investigation. \\
If variability studies sre performed on wide field of view cameras, the probability that an artificial variation of luminosity induced by a satellite occultation increase accordingly with the number of satellites in the FOV; the luminous variation timing will help to discriminate between natural and artificial variability.

\section{Radio Astronomy, how much is affected?}
Even with the best coating and mitigation procedures to decrease the impact on visual astronomical observations, what is often omitted, or forgotten, is that telecommunication constellations will shine in the radio wavelength bands, observable from the ground, not only from reflection of solar radio flux, but above all because of the need to broadcast internet (and other TLC services) from the satellite networks to the ground stations.\\

Radio astronomers have been engaged for decades in the work of the
United Nations Agency ITU to regulate the international use of the
radio frequency spectrum. Their efforts ensured a limited number of
narrow bands of the spectrum received protection to allow radio astronomy
to develop and conduct essential and unique research.

Despite the special international protection for radio-astronomy,
some sources of radio frequency interference (RFI) are inescapable.
While radio astronomers can minimize the effect of many terrestrial sources by placing their telescopes in remote sites, none can escape from RFI generated by satellites' constellations, because the reason for existence of these networks is to provide ubiquitous TLC signals even in the most remote part of the globe.\\

Very often in place where radio observatories are placed (e.g. at the two SKA sites in Australia and South Africa) there is legislation to
protect the telescopes from ground-based radio interference at those
frequencies, the use of air and space-borne radio communications is
regulated on a collaborative international basis.

So this development of TLC satellites' constellations are not intended as a collaborative effort, but as a \underline{unilateral commercially-driven private need}.\\

The scientific needs of radio astronomers and other users of the passive services for the allocation of frequencies were first stated at the World Administrative Radio Conference held in 1959 (WARC-59). At that time, the general pattern of a frequency-allocation scheme was:\\
\begin{enumerate}
\item that the science of radio astronomy should be recognized as a service in the Radio Regulations of the \textbf{International Telecommunication Union (ITU)}; 
\item that \textbf{a series of bands of frequencies should be set aside
}internationally for radio astronomy\footnote{These bands should lie at approximately every octave above 30 MHz and should have bandwidths of about 1 percent of the center frequency.}
\item that special international protection should be afforded to the \textbf{hydrogen
line} (1400-1427 MHz), the \textbf{hydroxyl (OH) lines} (1645-1675
MHz), and to the predicted \textbf{deuterium line} (322-329 MHz)...
\end{enumerate}

Since 1959 a large number of spectral lines from a wide variety of
atoms and molecules in space have been discovered; the frequency
range of radio astronomy now extends to at least 500 GHz. In particular,
frequencies of the CO molecule (at 115, 230, and 345 GHz), isotopes
(at 110, 220, and 330 GHz) and the maser of H$_2$O at 22.235 GHz {[}29{]},
are critical to many aspects of astronomy, see also {[}25{]}. \\
\subsection{Which impacts on bands used for radio-astronomy?}

There are lot of professional on-ground radio astronomy facilities and very few radio bands free from telecommunications uses, see Fig. 12. Satellite constellations described in Table 2 use frequencies in the following bands:  \underline{L, S, C, X, K$_u$, K$_a$ and V.} \textit{Only K, W, D, G and Y bands are free from satellites uses.}\\
It is possible to fill a list of radio telescopes and projects affected by radio interference coming from the satellites' broadcast of internet/phone signals, see Table 3. \\
\begin{table}[htbp]
\centering
\caption{\bf Satellite-Constellations interference with ground-based radio astronomy observatories}
\begin{tabular}{|c|c|c|c|}
\hline 
\textbf{Radio Project} & \multicolumn1{p{25mm}}{\textbf{Constellations Name}} &  \multicolumn1{p{15mm}}{\textbf{Interference Bands}} & \textbf{No of Sats}\tabularnewline
\hline 
\hline 
\textbf{ASKAP} &  \multicolumn1{p{25mm}}{Iridium, Roscosmos, Chinese, SkyAndSpace, CASC} &  \multicolumn1{p{15mm}}{L} & 1520 \tabularnewline
\hline 
\textbf{SRT+I-VLBI }&  \multicolumn1{p{25mm}}{Iridium, Roscosmos, Chinese, SkyAndSpace, Xinwei, Astrotech} &  \multicolumn1{p{15mm}}{L, C, K} & 5082 \tabularnewline
\hline 
\textbf{MeerKAT} & \multicolumn1{p{25mm}}{Iridium, Roscosmos, Chinese, SkyAndSpace, Globalstar, OneWeb, SpaceX, Kepler, LuckyStar, Xinwei, Astrotech} &  \multicolumn1{p{15mm}}{S, C, X, K$_u$} & 17288 \tabularnewline
\hline 
\textbf{SKA1 }&  \multicolumn1{p{25mm}}{Iridium, Roscosmos, Chinese, SkyAndSpace, Globalstar, OneWeb, SpaceX, Kepler, LuckyStar, Xinwei, Astrotech} &  \multicolumn1{p{15mm}}{S, C, X, K$_u$} & 2038 \tabularnewline
\hline 
\textbf{VLA} &  \multicolumn1{p{25mm}}{Iridium, Roscosmos, Chinese, SkyAndSpace, Globalstar, Telesat, SpaceX, LeoSat, Boeing, SES, ViaSat, Karousel, Luckystar, Xinwei, Astrotech} &  \multicolumn1{p{15mm}}{S, C, X, K$_u$, K$_a$} & 21788 \tabularnewline
\hline 
\textbf{ngVLA} &  \multicolumn1{p{25mm}}{Iridium, Roscosmos, Chinese, SkyAndSpace, Globalstar, Telesat, SpaceX, LeoSat, Boeing, SES, ViaSat, Karousel, CASC, LuckyStar, Xinwei, Astrotech} &  \multicolumn1{p{15mm}}{S, C, X, K$_u$, K$_a$, V} & 48108 \tabularnewline
\hline 
\textbf{ngVLA-LBA} &  \multicolumn1{p{25mm}}{Iridium, Roscosmos, Chinese, SkyAndSpace, Globalstar, Telesat, SpaceX, LeoSat, Boeing, SES, ViaSat, Karousel, CASC, LuckyStar, Xinwei, Astrotech} &  \multicolumn1{p{15mm}}{S, C, X, K$_u$, K$_a$, V} & 48108 \tabularnewline
\hline 
\end{tabular}
\end{table}

The development of the latest generation telecommunication networks (both from space and from Earth) already has a profound impact on radio-astronomical observations (at all sub-bands): with LEO satellite fleets it is quite certain that the situation could become unbearable.\\

\begin{figure}[t]
\centering
\fbox{\includegraphics[width=0.9\linewidth]{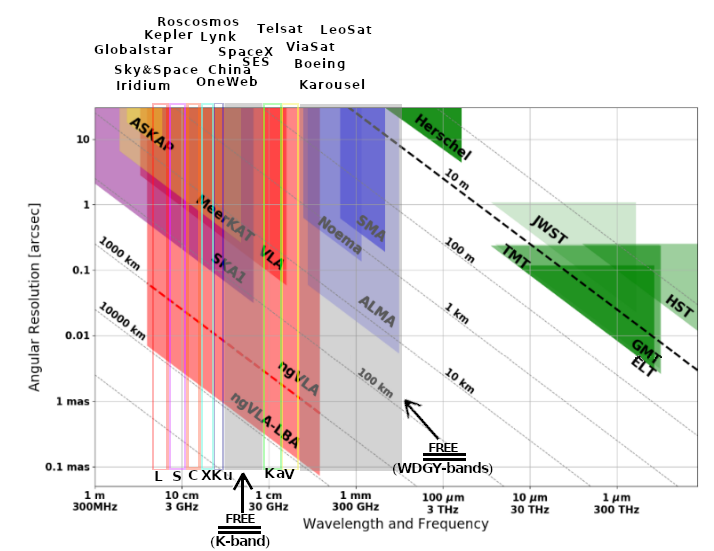}}
\caption{ Spatial resolution versus frequency set by the maximum baseline of the ngVLA compared to other existing and planned facilities. }
\label{fig:radiocomparisons}
\end{figure}

In particular, low Earth orbit satellite\textquoteright s spectral
windows identified to communicate with Earth stations in the L (1-2 GHz), S (2-4 GHz), C (4-8.2 GHz), X (8.2-12.5 GHz), K$_u$ (12.5-18 GHz), K$_a$ (27-40 GHz) and V (40-75 GHz) bands will overlap with the nominal
radio-astronomy bands, so will interfere with ground based radio telescopes
and radio interferometers, making the radio detectors enter in a non-linear regime in the K band (18-26.5 GHz) and Q band (33-50 GHz). This fact will irreparably compromise the whole chain of analysis in those bands, with repercussions on our understanding of the Universe, or even possibly making the astrophysics community blind to these spectral windows
from the ground, even if they are formally free from telecommunications broadcast, see Fig. 12.\\

From Fig. 12 it is possible to identify free bands from TLC uses, which are: K, W, D, G and Y. Although they are theoretically free from uses, it has been noted that TLC providers usually do not respect the authorized frequency windows and often produce unauthorized RFI interference in bands formally free (e.g. K-band, as experimented in SRT, where at "some" satellite transit the SRT K-band detector saturates, entering into a non-linear regime).  

There are different projects in development for ground-based radio-astronomy that will significantly overlap with telecommunication signals coming from the satellites' constellations in orbit, see Table 3 for details: 
\begin{itemize}
\item \textbf{Australian Square Kilometre Array Pathfinder, ASKAP} see {[}46{]}: located in Australia. The ASKAP will use 4 radio-bands: 0.7-1.012 GHz, 0.9-1.2 GHz, 1.2-1.52 GHz, 1.48-1.78 GHz. 
\item \textbf{Sardinia Radio Telescope, SRT} and \textbf{SRT
+ Italian Very Long Baseline Array, SRT I-VLBA} see {[}40{]}: located in Sardinia, Italy. The SRT will use 4 radio-bands: 0.3-0.4 GHz, 1.3-1.86 GHz, 5.7-7.7 GHz, 18-26.5 GHz. 
\item \textbf{South African MeerKAT radio telescope, MeerKAT} see {[}45{]}: located in Northern Cape, South Africa, is the precursor of SKA. MeerKAT operates in the L-Band at frequencies 0.9-1.67 GHz and UHF-Band between 0.5-1.015 GHz. 
\item \textbf{Next Generation Very Large Array, ngVLA} and \textbf{ngVLA
Long Baseline Array, LBA} see {[}22{]}: located in New Mexico, west Texas,
Arizona, and northern Mexico. The ngVLA will use 6 radio-bands: 2,4 GHz,
8 GHz, 16 GHz, 27 GHz, 41 GHz and 93 GHz. 
\item \textbf{Square Kilometer Array, SKA} see {[}23{]}, {[}27{]} will interfere with K$_u$ communication bands.
\item \textbf{Atacama Large Millimeter Array, ALMA} see {[}26{]}, the world-leading mm and sub-mm observatory built in Atacama, Chile with enormous sums spent by a broad international community, is a facility that has brought us many significant discoveries and played a crucial role in the global system of EHT (first image of ever of a black hole, published in April 2019), has its Bands 1, and 2+3 exactly in the potentially polluted part of the spectrum, WDGY bands, not part of 5G satellite communications, but foreseen for 6G technology in few years. 
\end{itemize}

\begin{figure*}[t]
\centering
\fbox{\includegraphics[width=0.9\linewidth]{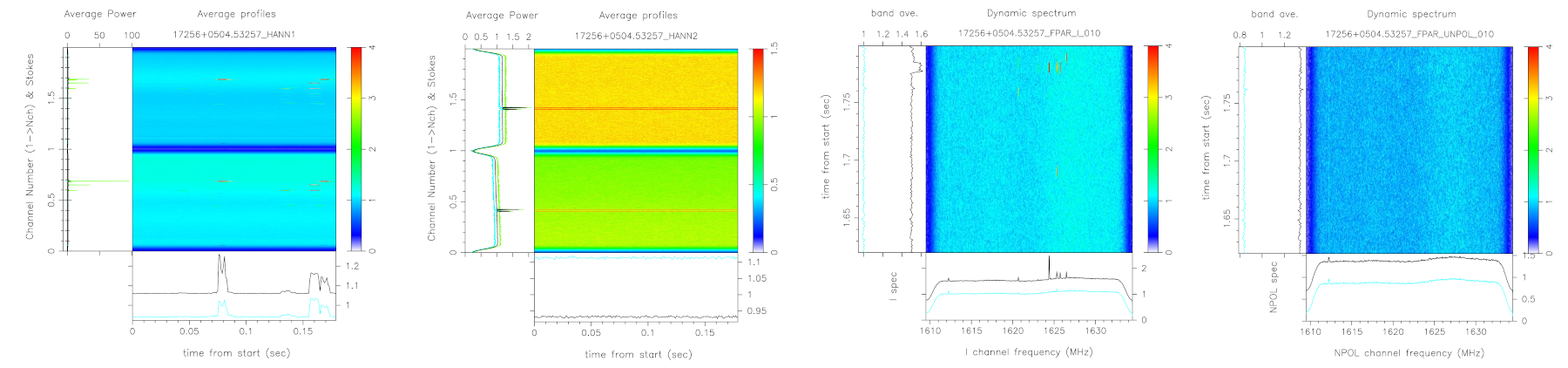}}
\caption{ In the first image, a typical 180 ms long dynamic spectra for the dual linear polarization channels (X and Y, in the upper and lower halves of the main panel), are shown, each observed across a 25 MHz wide band centered at 1622 MHz. The two line plots in the bottom panel show the corresponding time profiles of the band-averaged intensities in X and Y, respectively. In the second image near a bandwidth of 3.125 MHz in the RAS band (centered at 1612.5 MHz), which is narrow and so has a correspondingly better spectral resolution, providing a more resolved view of the line features from the star. The third image is a 180 ms sequence of Stokes I spectra taken at 1 ms intervals, together with their arithmetic (black) and robust(coloured) means on each axis, so that only the \(\sim\, 1612\) MHz feature comes from the star. The last image shows a dynamic spectrum with the unpolarized component \(Iu\) (i.e. after removing the polarized contribution from the Stokes I), and thus mostly free of RFI. }
\label{fig:iridium}
\end{figure*}

To aggravate the matter, with the current technological development,
the planned density of radio frequency transmitters is impossible
to envisage. In addition to millions of new commercial wireless hot
spot base stations on Earth directly connected to the approx. 50,000
new satellites in space, we will produce at least 200 billion new
transmitting objects, according to estimates, as part of the Internet
of Things (IoT) by 2020-2022, and \textit{one trillion} objects a few years later. \\

Such a large number of radio-emitting objects could make radio astronomy
from ground stations impossible without a concentrated protection effort made by countries\textquoteright{} safe zones where radio astronomy facilities are placed. \\
\textit{This should be followed by an international moratorium to limit satellites' communications emissions.}\\

\subsection{Mitigate TLC interference: the Iridium case study}

This analysis is based on Avinash A. et al 2019, see [24]. The International Telegraphic Union (ITU) granted the Radio Astronomy Service (RAS) primary status in the 1610.6-1613.8 MHz band in 1992 to observe the 1612.235 MHz spectral line emission from the hydroxyl molecule (OH). This is typically emitted by OH/IR stars (see Lewis, Eder \& Terzian 1985) as a pair of narrow features, with the allocated band sized to allow for Doppler shifts of the emission, as well as guard-band separation from ITU Services using an adjacent spectrum. One such is the Iridium L-Band system, which presently uses a 1618.85-1626.5 MHz allocation.\\

But this system also produces a comb of RFI, with a characteristic 333 kHz spacing, extending well beyond its licensed band. The \(\sim1\, Jy\) intensity of this comb in the RAS band for most of observatories is comparable with the signal from many of the brighter OH/IR stars. No pre-launch simulation available to the ITU or to radio-astronomers gave any hint of the existence of this noxious artifact when the System was granted spectrum. Hence the need now for RFI mitigation. \\

The study was made using the highly sensitive 305 m Arecibo telescope, which has 80 dB of forward gain and thus a narrow main beam. Accordingly the 66 active, low-earth-orbit satellites from the Iridium System are only generally seen at Arecibo in distant side lobes: that lessens their impact on our observations.\\

Iridium uses a frequency multiplexed – time multiplexed operational mode. This gives two helpful features, as the signal is strongly polarized (more specifically, has right hand circular polarization), and more unusually, has a satellite down-link time multiplexed in exactly the same band as the phone hand setup-link signal.\\

Each Iridium satellite operates on a 90 ms cycle, with half assigned to the up-link, and half to the down-link. That allows the folding of data at a secondary period of 180 ms, or twice the basic cycle. The timing operations of the entire Iridium constellation are governed by the System’s most intense signal, the 1626 MHz clock synchronization signal.\\

Data were acquired with full Stokes \((I, Q, U, V)\) parameters as high time-resolution, single dish spectra. Dynamic 1024 channel spectra were recorded every millisecond using the 9 level sampling mode of an auto-correlator simultaneously in both the Iridium band, using a 25 MHz bandwidth centered at \(\sim\,1622 MHz\) and in the RAS band using a 3.125 MHz bandwidth with proportionately finer spectral resolution. These accumulate the net temporal correlation between fluctuations in every possible combination of spectral channels from every possible pairing of spectra: they are produced as cross-correlations between spectra in the native linears, in I, in the unpolarized flux,  \(Iu=I-\sqrt{Q2+U2+V2}\) , as well as between the two observed bandwidths and as auto-correlations of spectra. After folding 2 minutes of data at the period of the Iridium clock cycle, the only Iridium artifacts in our RAS band data are momentary gain compression episodes.\\

The bandpass gain calibration is applied to the linear polarization data before computing Q \& Iu from the Stokes I, Q, U \& V of each channel of each spectrum as  \(Iu=I-Ip\)  (where the polarized component is estimated as \(Ip=\sqrt{Q2+U2+V2}\), which reduces the residuals from Iu.

Fig. 13 shows a typical 180 ms (two-cycle) sequence of Stokes I spectra with the OH/IR star at \(\sim\,1612\) MHz and Iridium's signal between 1620-1627 MHz. The mean side-lobe response of the strongest Iridium Stokes I feature here, which is only seen for \(\sim\,5\%\) of the time, has \(\sim\,7\) times the intensity of strong OH/IR star which is seen with the main beam. So Iridium can easily saturate an astronomical receiver. \\

\subsection{How to extend the case study to other bands?}

Mitigation techniques can help to reduce RFI significantly up to tens of DB, but never reduced to zero. A more detailed approach to different transmitters coming from other satellites needs a special study on the nature and modulations of those signals in order to apply such techniques. \\

Some radio astronomy sites (e.g. NRAO) rely on Radio Quiet Zones, but it is not possible to prevent satellites from shining on the ground. So SpaceX has promised to accommodate NRAO by not emitting in the direction of the Green Bank Observatory. This will be easy because some satellites use \underline{phased array technology}, aiming beams only in the direction of user devices. Since there will be no user devices in the Radio Quiet Zone, there will not be any interference from SpaceX with the NRAO telescope. This will also be true of other large constellations that use millimeter waves and phased array technology. In particular, the phased array technology, see [59], can be used with transmitting devices operating above 20GHz, so with the K$_a$, Q, U, E, V, W, D \& G - bands; from Table 2 it is possible to find satellite constellations with beam forming capabilities: SpaceX, Telesat, LeoSat, Chinese Aerospace, Boeing, SES, ViaSat, Karousel, CASC and Xinwei leaving out OneWeb, Iridium, Roscosmos, Sky and Space Global, Globalstar, SatRevolution, LuckyStar, Commsat and AstroTech, equipped with classical radiation cones.\\ 

Even if a phased array can be used to exclude some places from satellite signals, this mitigation can not be considered as conclusive, because 1) the phased array does not emit zero signal outside the principal phased direction, but only produces disruptive interference to guarantee only a principal direction of propagation, but outside the main beam the signal is not zero as manufacturers declare, so filling out the whole sky with 50 thousand satellites will increment the Night Sky Background in each radio-band; and 2) the possibility to avoid signals in Radio Quiet Zones assumes that no user will ask for the service, neither single users nor passing users (from ships, cars, planes, trains, etc.). Finally, 3) because the single radio observatory can work in conjunction with other partners all around the world (in the Very Long Baseline Interferometry capability), it is required that each observatory of the long baseline to be free of pollution in the desired receiver band. This is not possible at all.\\

The principal issue related to the RFI induced by the satellites' constellations is that each ground-based astronomical receiver needs to discriminate between each satellite transmitter band, so astronomical detectors will find in their data many different RFI coming from different satellites, each one with a different polarization and timing. The mitigation process could be very difficult, if not impossible, without a good characterization of each satellite signal. \\

With all these orbiting satellites, if in theory this cleaning process is possible, in practice the work of extracting reliable scientific content from radio-astronomy data could be so tricky that the overhead needed in terms of computation time and computing teraflops (e.g. electrical power consumption) will explode. This shows that a clear evaluation and quantification of the damage is not simple like the optical investigations. \\ 

The only working mitigation possibility is to absolutely preclude some bands from telecommunications and use them (only them) for radio-astronomy; clearly surrendering all opportunities to perform a complete multi-wavelength analysis on astronomical objects.    

\section{Cherenkov Astronomy: Are Telescopes and Arrays affected?}

Gamma-ray photons have energies (E>125 keV) that span many orders of magnitude, from MeV to TeV and beyond. Consequently a single detector technology will not be able to cover the entire gamma-ray range.\\

While gamma rays with energies on the order of MeV and GeV are detected by satellite instruments, \underline{very high energy} (VHE, E>50 GeV) gamma rays can be efficiently detected, with large collecting areas, only from the ground, e.g. with arrays of Cherenkov Telescopes (see Fig. 14).
\\
\begin{figure}[t]
\centering
\fbox{\includegraphics[width=0.9\linewidth]{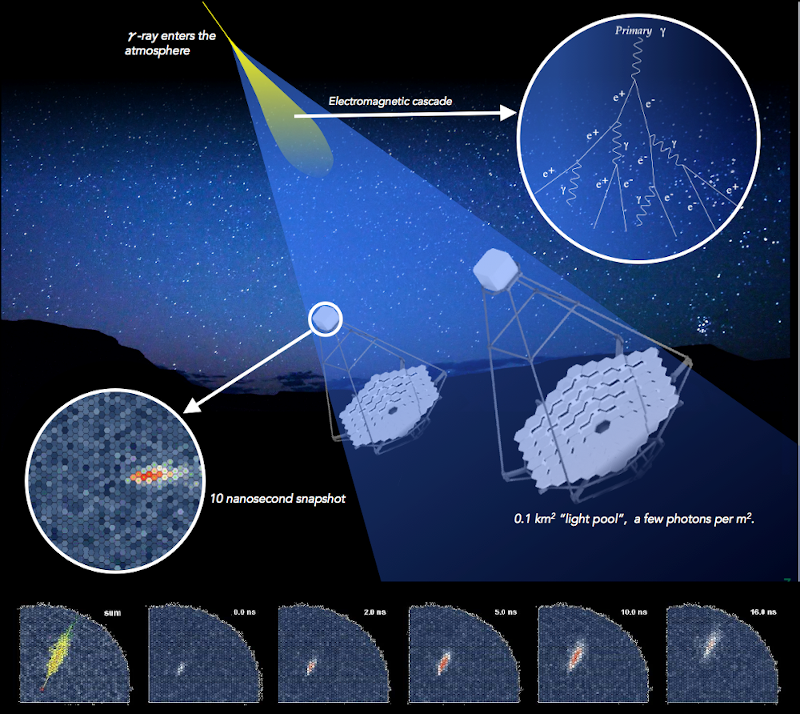}}
\caption{ The picture clearly shows the time evolution of a shower originated by a 10TeV gamma ray hit at 250m with a time-sampling of \(\sim\,1\,ns\). The original gamma direction can be reconstructed using a special parametrization, and the use of two (or more) different detectors simultaneously can cut-off all spurious events. }
\label{fig:showers}
\end{figure}

Depending on the energy of the initial VHE cosmic gamma ray, there are many electron/positron pairs in the resulting cascade that are capable of emitting Cerenkov radiation.\\

As a result, a large pool of Cerenkov light comes with the air shower particles. The Cherenkov emission is at the ultraviolet and visible wavelengths, and the shower can be imaged directly with an array of pixels, within a camera placed at the telescope's focus. \\

Thus for Cherenkov telescopes the atmosphere has to be considered as part of the "detector," and the air shower analysis permits counting high energy photons, their origin and their spectral signature (e.g. their energy).

Can these kinds of ground-based facilities suffer from light pollution coming from satellites’ constellations?\\

Considering that the data reduction and analysis strategies are totally different from the optical case, to inspect the possible damages due to the satellites' constellations it is possible to start analysing the FOV of some telescope involved in Cherenkov observations.\\

\begin{table*}[t]
\centering
\fbox{\includegraphics[width=0.9\linewidth]{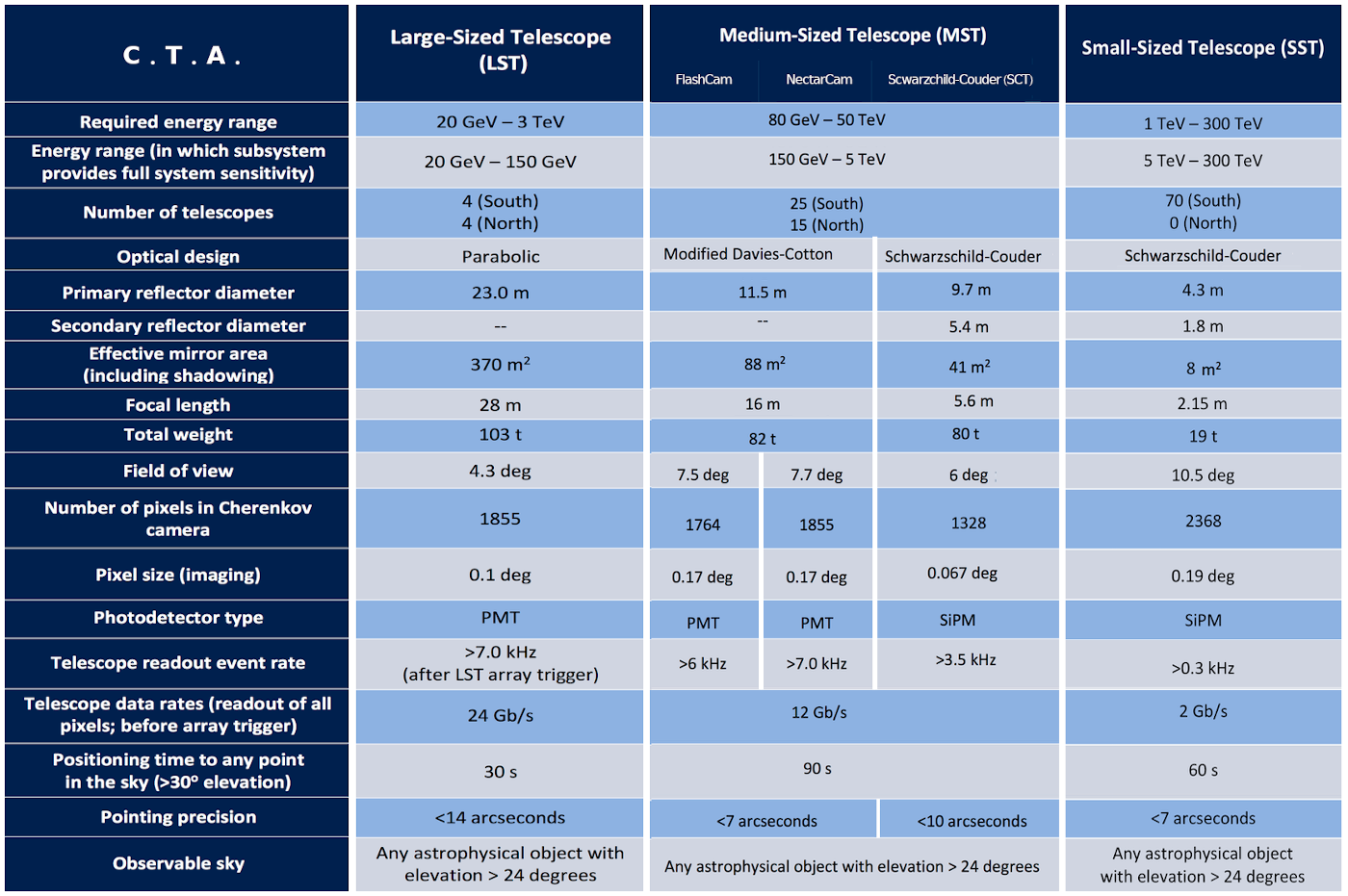}}
\caption{ In the table are represented three different kinds of Cherenkov telescopes within CTA arrays and their main characteristics. }
\label{tab:tabellacta}
\end{table*}

There are different facilities for Cherenkov gamma-ray astronomy in operation and development. The current-generation of Cherenkov telescopes comprises the ASTRI \& mini-Array project, see [47]; H.E.S.S., see [50]; the MAGIC array, see [48] and VERITAS array, see [51]. The most important next-generation facility will be the the Cherenkov Telescope Array (CTA), see [49].\\

CTA will be composed of two arrays of Cherenkov telescopes of different sizes (large-sized telescopes, LST \( \sim\,24\,m\); medium-sized telescopes, MST, \( \sim\,12\,m\); small-sized telescopes, SST, \( \sim\,4\,m\) that will be placed in two different sites, see Table 4:

\begin{itemize}
\item \(\sim20\) telescopes for the \underline{CTA North-site} (Canary Islands, Spain)
\item \(\sim100\) telescopes for the \underline{CTA South-site} (Chile)
\end{itemize}

While the CTA array in the northern hemisphere will be more limited in the number of telescopes, and will be more focused on the low and middle VHE range (from ~ 20 GeV to ~ 20 TeV), the CTA array in the southern hemisphere, with its prime view of the rich central region of our Galaxy, will span the entire energy range accessible to CTA, covering gamma-ray energies from 20 GeV to 300 TeV.\\

The three classes of telescopes will be distributed in the two array sites, with possible layout configurations as depicted in Fig. 15.

\begin{itemize}
\item SST \(\oslash \sim 4\,m\ \rightarrow\, E>10\,TeV\rightarrow\,FOV\sim10\,deg\) 
\item MST \(\oslash \sim 12\,m\ \rightarrow\, E\sim [0.1;\,1]\,TeV \rightarrow\,FOV\sim 6-8\,deg\) 
\item LST \(\oslash \sim 24\,m\ \rightarrow\, E<\, 0.1\,TeV \rightarrow\,FOV\sim 4-5\,deg\)
\item ASTRI/MA \(\oslash \sim 4.5\,m\ \rightarrow\, E>\, 1\,TeV \rightarrow\,FOV\sim 9.6\,deg\)
\item MAGIC \(\oslash \sim 17\,m\ \rightarrow\, E<\, 1\,TeV \rightarrow\,FOV\sim 3.5\,deg\)
\item VERITAS \(\oslash \sim 12\,m\ \rightarrow\, E\sim [0.1;\,30]\,TeV  \rightarrow\,FOV\sim 3.5\,deg\)
\item H.E.S.S. \(\oslash \sim 28\,m\ \rightarrow\, E<\, 20\,TeV \rightarrow\,FOV\sim 5\,deg\)
\end{itemize}

\begin{figure}[htbp]
\centering
\fbox{\includegraphics[width=0.9\linewidth]{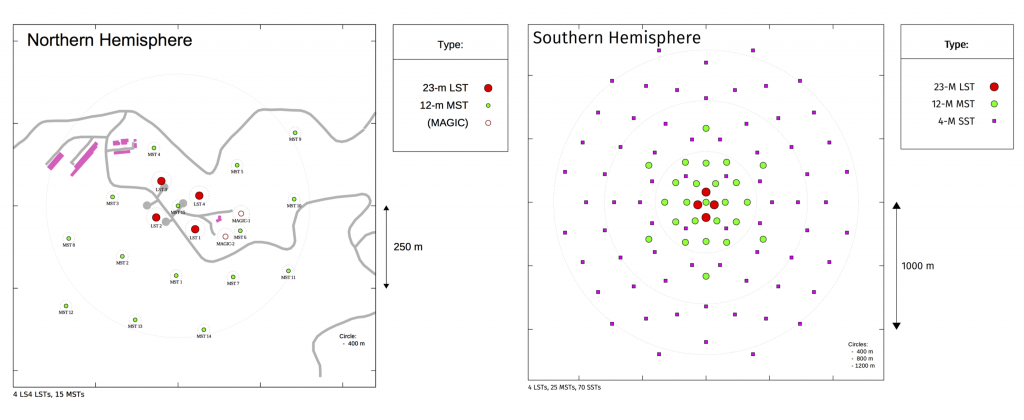}}
\caption{A potential layout of the telescope arrays in both the northern and southern hemispheres.}
\label{fig:ctaarrays}
\end{figure}
\subsection{Quantifying Cherenkov damages}
Considering a mean exposure time of about 20 minutes, the total number of satellites in the FOV will depend on the FOV area of each telescope. However, in order to estimate the rate of spurious
triggered events, it is necessary to compute the single camera pixel crossing time (i.e. the mean time that any satellite will remain within a single pixel), which is for a typical SST camera of the order of \(\sim2-3\,sec\).\\

Taking into account equation (1), (2), (3) and Table 2 within the FOV of Cherenkov telescopes of similar characteristics of those used in CTA, the crossing time is:\\
\\
\begin{tabular}{|c|c|c|c|c|}
\hline 
\multicolumn1{p{10mm}}{\textbf{FOV [deg]}} & \multicolumn1{p{10mm}}{\textbf{Sats per FOV}} & \multicolumn1{p{10mm}}{\textbf{t.cross @ 340Km}} & \multicolumn1{p{10mm}}{\textbf{t.cross @ 550Km}} & \multicolumn1{p{10mm}}{\textbf{t.cross @ 1150Km}} \tabularnewline
\hline 
\hline 
$LST\rightarrow 4.3$ & 14 & 65[s] & 68[s] & 78[s] \tabularnewline
\hline 
$MST_1\rightarrow 6.0$ & 28 & 91[s] & 95[s] & 109[s] \tabularnewline
\hline 
$MST_2\rightarrow 7.5$ & 44 & 113[s] & 119[s] & 136[s] \tabularnewline
\hline 
$MST_3\rightarrow 7.7$ & 46 & 116[s] & 122[s] & 140[s] \tabularnewline
\hline 
$SST\rightarrow 10.5$ & 86 & 159[s] & 166[s] & 190[s] \tabularnewline
\hline 
\hline 
$_{ASTRI}\rightarrow 9.6$ & 79 & 146[s] & 153[s] & 173[s] \tabularnewline
\hline 
$_{MAGIC}\rightarrow 3.5$ & 11 & 52[s] & 55[s] & 63[s] \tabularnewline
\hline 
$_{VERITAS}\rightarrow 3.5$ & 11 & 52[s] & 55[s] & 63[s] \tabularnewline
\hline 
$_{H.E.E.S.}\rightarrow 5$ & 23 & 75[s] & 79[s] & 90[s] \tabularnewline
\hline 
\end{tabular}\\

The total number of satellites crossing the FOV of a \underline{typical cherenkov 20-minutes exposure} are respectively:\\
\begin{tabular}{|c|c|c|}
\hline 
    \(\sim\,30\,for\,LST\) & \(\sim\,40\,for\,MST_1\) & \(\sim\,54\,for\,MST_2\)\tabularnewline 
\hline 
    \(\sim\,56\,for\,MST_3\) & \(\sim\,90\,for\,SST\) & \(\sim\,86\,for\,ASTRI\) \tabularnewline
\hline 
    \(\sim\,32\,for\,MAGIC\) & \(\sim\,24\,for\,VERITAS\) & \(\sim\,25\,for\,H.E.S.S.\) \tabularnewline
\hline 
\end{tabular}
\\
This time could be enough to trigger a spurious event, but depends on the brightness of the satellites.\\

If a star of 3rd magnitude is considered (e.g. Ztauri is 2.97 mag), it can trigger a Cherenkov camera, under particular trigger configurations.\\

During observations near twilight conditions, the 2nd and 3rd magnitude satellites in FOV could impose a different trigger topological strategy and higher single pixel trigger threshold, which would result in an higher energy threshold of the detector. In fact, in the case of an almost-perfect alignment, a 5NN topological logic should avoid (almost completely) any possible spurious trigger due to bright stars or passing (point-like) artificial objects.\\

In addition to this, an individual pixel rate (IPR) control should help in order to be sure not to get any spurious trigger.

\begin{figure}[htbp]
\centering
\fbox{\includegraphics[width=0.9\linewidth]{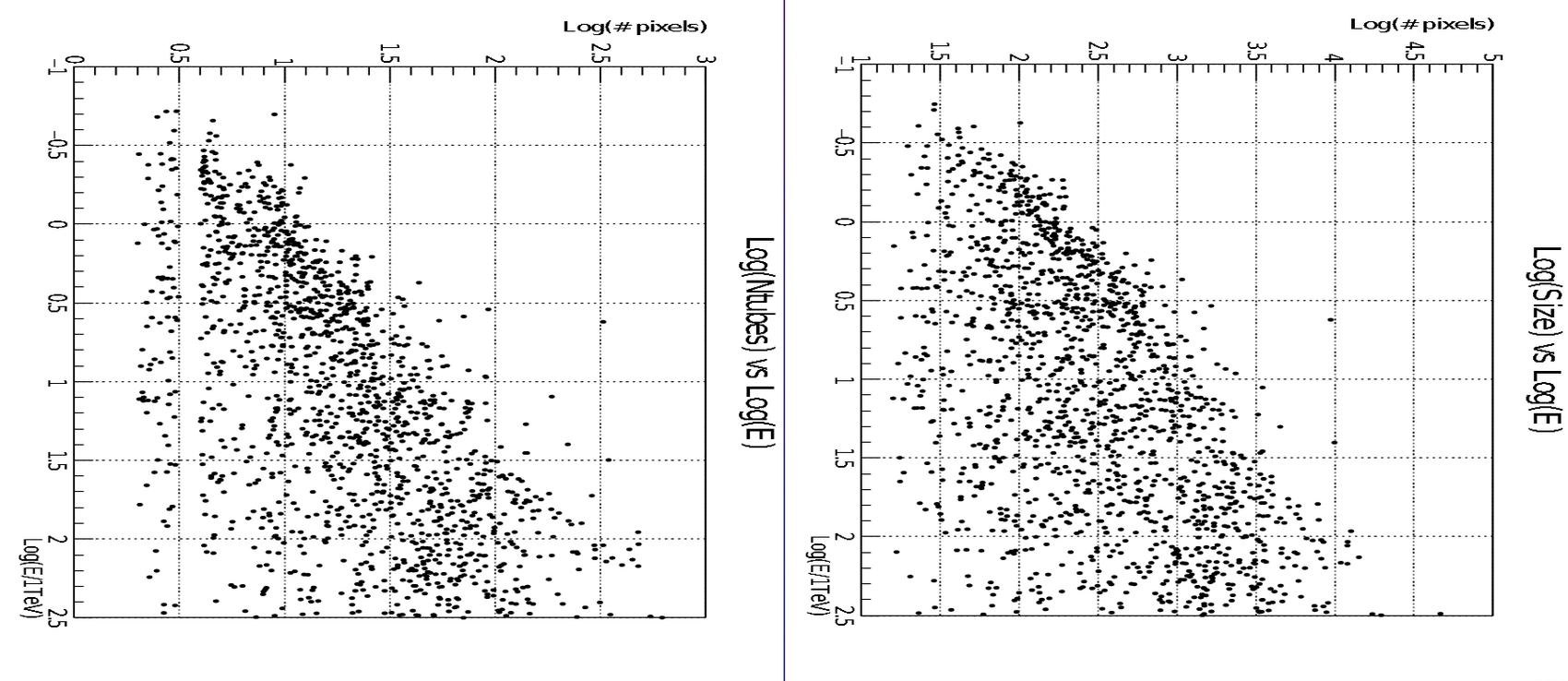}}
\caption{Even at the highest energy we simulated so far the shower image can be quite small, both in term of number of pixels and size, credits C. Bigongiari.}
\label{fig:astrigraph}
\end{figure}
On the contrary, satellites with higher magnitude cannot trigger the telescopes. Nevertheless, they can increase the background light on a certain number of pixels (depending on the PSF and NSB).\\

For instance, satellites with magnitude 7th will more than double the single pixel background light, while satellites with magnitude 5th will add a factor of  \(\sim\, 7\) more background light.\\

Pixels with higher levels of background light could degrade the reconstruction of Cherenkov images; in such cases it would not even be possible to remove the contribution of the affected pixels by increasing the levels of the image cleaning procedures.\\

In principle smaller images, which are not necessarily due to low energy events, are the most affected by this spurious effect.\\

Since the additional light hits (in principle) single pixels (within a given Cherenkov event trigger time window), the cleaning levels can likely be the default ones, providing information on the "hot" pixels (for each pixel and each Cherenkov trigger) is made available.\\

To do so, the level of the pixels' NSB should be monitored (e.g. with dedicated quality monitoring data) with a rate higher than the inverse of the time needed for the artificial object to cross
a single pixel (\(\sim\, 3\, sec\)).

\subsection{Remarks in Cherenkov Astronomy Damages}

In conclusion, the more affected cherenkov telescopes will be those at higher energy ranges, i.e. SST-like telescopes; MST \& LST-like telescopes will be less affected.\\ 

The following list shows all concerns arisen for SST-like telescopes:
\begin{itemize}
\item Lowest orbit satellites may indeed provide additional non-negligible source of NSB in the Cherenkov cameras.
\item Higher orbit satellites might have some impact on the rate of single pixels; in order to be \(90\%\) sure to avoid spurious triggers, a suitable trigger logic with a >=4NN topology may be indeed needed, while an IPR control may be also implemented.
\item In order to always be able to use optimal image cleaning levels, the pixels hit by the artificial light must be somehow suppressed. Thus, the information of the NSB amount for each cameras' pixel should be monitored by means of dedicated quality monitoring data, in time steps smaller than the typical time in which the artificial objects cross a single pixel (\(\sim 3sec\))
\end{itemize}

\section{Short-term and Long-term Impacts from Mega-constellation Satellites, and the Future of the Dark Sky}
\begin{figure}[htbp]
\centering
\fbox{\includegraphics[width=0.9\linewidth]{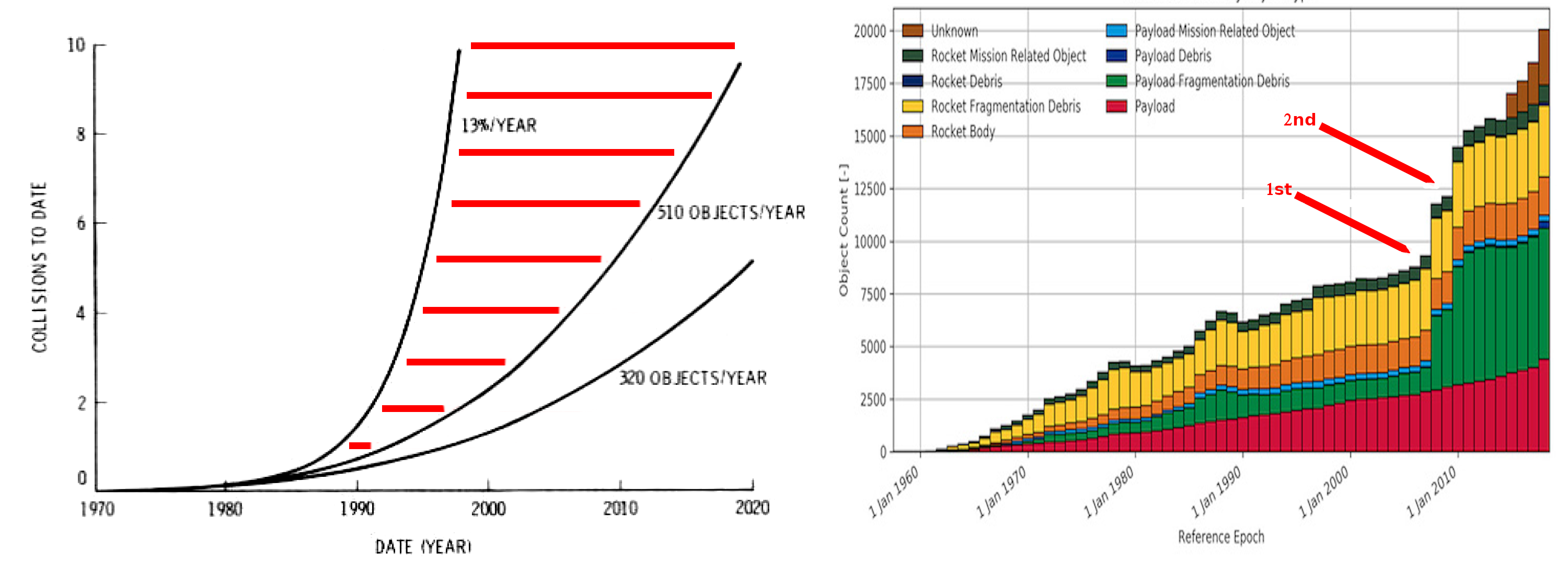}}
\caption{Respect to Kesslers' computation and prediction of the total collisions by the given date under various growth assumptions in the left picture, in 1978 the first collision was expected between 1989 and 1997, and happened around 2008 and 2010, see right picture. In each collision the number of orbiting debrids increase dramatically.}
\label{fig:kess00}
\end{figure}

Space debris designates any artificial object orbiting the Earth no longer serving its initial function (i.e. fragmented spacecraft parts, abandoned launch vehicle stages, disposed spacecraft, etc.). Currently, there are over \underline{500,000} pieces of space debris the size of a marble or larger orbiting the Earth, traveling up to 17,500 mph. Millions of others are currently untraceable. In addition, around \underline{4,000 active and inactive satellites} are presently in Earth’s orbit. As early as 1978, Donald Kessler had published a paper, see [34], detailing and warning us of the devastating cascading effect of collision-induced debris creation (the so-called Kessler syndrome, see further below).\\
Kessler wrote that the finite probability of satellites colliding and creating more debris fragments, would thus create a snowball effect which, were the tipping point to be reached, would cause devastating consequences not only for the future of space exploration, but also to our environment, see Fig. 18. \\

\begin{figure}[htbp]
\centering
\fbox{\includegraphics[width=0.9\linewidth]{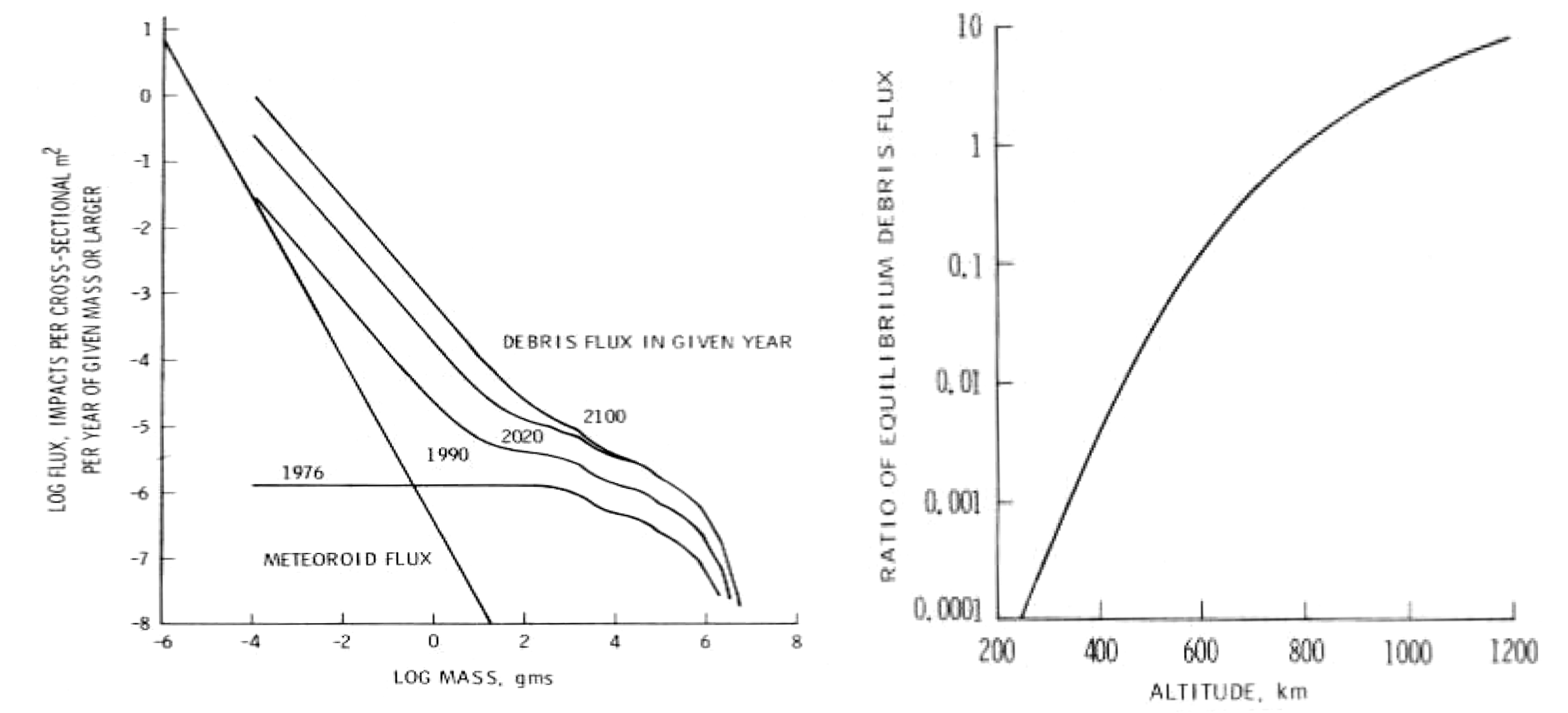}}
\caption{Average debris flux between 700 and 1200km altitude, assuming a net input rate of 510 satellites per year and no atmospheric drag. With satellites' mega-constellations there will be at least 10 times the input rate in LEO orbit. }
\label{fig:kess0}
\end{figure}
The Kessler paper simulates the creation of a debris belt around the Earth following an increase of orbiting satellite numbers with a rate of 510 satellites per year (expecting a satellite rate launch reduced to zero in 2020): with mega-constellations this rate is instead ten times that expected by Kessler. Just considering SpaceX, it has scheduled a deorbiting rate with substitutions from 4,000 to 8,000 satellites per year.\\

While some measures, see [52], for the mitigation of space debris have been adopted over the years by the United Nations Office for Outer Space Affairs, the proposed seven guidelines are still rarely respected, see [53]. Furthermore, recent (successful) anti-satellite missile test attempts by China, see [54], and India, see [55], escalate the very real risk of the Kessler syndrome coming into effect much sooner than we think.\\

The Starlink mega-constellation satellite project, described previously in Section 2, is built and managed by SpaceX (currently paving the way for space commercial activities). It aims at deploying “the world’s most advanced broadband internet system” to “deliver high speed broadband internet to locations where access has been unreliable, expensive, or completely unavailable.” \\

As of October 2019, the U.S. Federal Communications Commission (FCC) had already approved 12,000 such satellites with another 30,000 pending, according to filings submitted by SpaceX to the International Telecommunication Union. A first large deployment of 60 satellites occurred in May 2019, quickly followed by three more in November 2019, January 2020 and February 2020. More recent and current launches have been occurring every month, while SpaceX announced that as many as 24 Starlink missions could be launched in 2020 alone (i.e. well over 1,000 satellites in one year alone). \\

This hastens an era of uncontrolled space pollution. The International Dark-sky Association, which advocates for the preservation and protection of clear dark skies, recently issued a statement, see [56], on the impact of mega-constellations, further raising concerns as to the prospect of such global-scale projects. \\

In its entirety, Starlink’s expected 42,000 satellites will not only be unprecedented, but also beyond any reasonable measure to adequately address the burning problem of low-Earth orbit pollution any longer (see also Sections 2 and 3). To put this into better context, nearly 9,000 satellites, probes and landers have been sent to space since 1957 – in other words, within the scope of just a few years, SpaceX (and readily every other major commercial constellation-satellite party, see Table 2) would launch considerably more satellites than humankind has ever launched in its entire history. In addition, the goals for this are embedded in timely economic and business prospects with a race to global 5G networks, prone to becoming a competitive market. By setting this precedent, SpaceX will certainly not be the sole purveyor of global internet access as part of this new mega-constellations market, see Table 2.\\

After the launch of a second batch of 60 satellites near the end of 2019, numerous disastrous events were reported by astronomers worldwide of Starlink satellites completely impairing telescope observations, see Fig. 2. \\

\begin{figure*}[t]
\centering
\fbox{\includegraphics[width=0.9\linewidth]{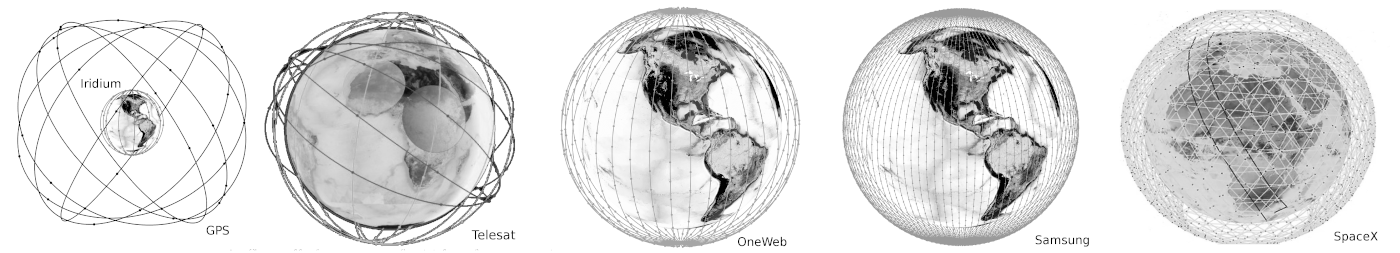}}
\caption{ In the four images are summarized different orbital configurations of a few satellites' constellations: GPS, Iridium, Telesat, OneWeb, Samsung and SpaceX. }
\label{fig:orbite}
\end{figure*}

These observatories have a record of producing state-of-the-art observations in some of the world’s most pristine night-sky locations, and the expansion of mega-constellation satellites will only further deteriorate, perhaps even completely undermine future observations. That only about \textbf{0.4\% of the planned 42,000 launched satellites} (omitting the grand total of over 60,000+ satellite-constellations planned or waiting for approval) is already causing distress and observable nocuous consequences should itself be cause for extreme warning in an already unfettered business market with little to no binding international regulations on the protection of the night sky. \\

Recent emergency maneuvers, see [57], had to be undertaken by the European Space Agency after SpaceX failed to respond on time, to avoid a near-collision with Starlink satellites (it was SpaceX’s responsibility to perform said maneuver). The recent key legal evidence that the FCC approval of the Starlink mega-constellation may have been unlawful, see [58], with regard to the National Environmental Policy Act (NEPA) which requires federal agencies to thoroughly assess the environmental impact of projects before accepting them, is furthermore attached to the scientific motives at the basis of the concerns in this manuscript. This case alone should thus be treated even more carefully as we advance into the uncharted territory of using space for such commercial activities.\\

Remarkable and inspiring night skies, which we have inherited to the pleasure of our eyes, hearts and imagination, are on the brink of being forever altered and desecrated more than they already are. On the basis of this analysis, we raise urgent concern with regard to the growing number of satellite constellations, and their impact not only on scientific observations, but also on space debris pollution and the preservation of the dark sky. International cooperation can take action by imposing thorough support and impact evaluations, in concert with international dialogue between regulatory agencies and satellite-constellation manufacturers. \\

Binding regulations and multi-party discussions among all and any current and future business enterprises by private entities are crucial for the preservation and protection of the night sky, and in preventing any potential disastrous Kessler syndrome effects. At present, these operations do not guarantee any safeguard for proper, reasonable or responsible commercial activity in Earth’s low-orbit. Legal and regulatory provisions along with moratoria on future launches should be imposed. The pursuit of satellite-constellation projects will most likely signal the end of humanity’s familiarity with space, scientific operations as detailed in this manuscript, and the dark sky as we know it.\\

In APPENDIX D is shown a legal approach on the possibility, under international laws, conventions and treaties to act towards the safeguard of the night sky by the whole international community.

\section{Conclusions: Quantifying Ground-based Astronomy Damages}

A large ground-based astronomical facility like the one taken as representative example, LBT, needs significant public funding to operate and give to the astronomical community scientific services and results. The  operations of the infrastructure are managed by the LBT Observatory organization, which shares operating costs among its partners.\\

It is not clear the total amount of cost for the infrastructure from start of construction to the current service operations, but it is possible to highlight the annual cost for the partnership: each partner spends approximately three billion euros per year. This implies a median cost for each observing night varying from 60,000 to 80,000 euros, depending on the seasons (length of the observing night).\\

If no good strategy to mitigate the flat-field pollution by satellites' trails is put into place, the scientific content of all images coming from the LBT astronomical observatory facilities will be strongly altered or compromised by an estimated 70-80\%. \\

Instead, if a good strategy on flat-fielding and calibration campaign can be found, the mitigation process could minimize this percentage to 30-40\% of the observing night. The value loss of the public investments committed to each astronomical on-ground facility due to damages from the satellites is proportional to the loss in the scientific content of related observations.\\

The LBT example is representative of other observatories and project partnerships (e.g. ESO VLT [7], Keck Observatory [11], etc.) because for each medium-to-large size optical observatory facility, the financial loss percentage could easily be projected and extended as a lower limit to large area survey observatories, leading to more adversely affected optical professional and amateur facilities. \\

A representative quantification for the Italian Institute of Astronomy cost and investments for astronomical ground-based projects \& facilities is \(>100 million euros\) for 2015-2017, see [44]:\\
\\
\begin{tabular}{|c|c|c}
\hline 
\multicolumn1{p{20mm}}{\textbf{\underline{Premiali}: CTA,SKA}} & \multicolumn1{p{20mm}}{\textbf{\underline{FOE}: CTA,SKA,ELT, SRT,LBT,TNG}} & TOT \tabularnewline
\hline 
\hline 
54,000,000 euros & 46,000,00 euros & 100,000,000 euros \tabularnewline
\hline 
\end{tabular}
\\

So each partner of any telescope community, because of the satellites' constellations, can quantify the related financial loss of value in principle, losing hundreds of millions of euros/dollars (of public money) per year, and cumulatively billions in decades.\\

For each nation and government, this financial loss has to be extended and integrated among each partnership in optical international collaboration projects.  \\
\textit{This is a very huge loss of value for the whole scientific community and for humankind as well}. \\ 

Remain totally unquantified the scientific economic loss for radio-astronomy as well as the loss of intangible asset of the whole humankind given by the loss of natural night sky.\\

Still unquantified is the possibility to enter in the Kessler syndrome with the creation of a very dense debrids belt around the Earth in low orbit; in particular the Kessler study underestimate the rate of satellites injection in orbit desiring the rate of launches coming near zero in 2020. Kessler did not find an exact number of satellites in orbit above which the probability of cascade impacts would reach 100\%, but it is possible that 50-100 thousand satellites in LEO could result in a very high probability of triggering a chain reaction in a very short time. This scenario will produce a unquantifiable damage to the whole Space esploration and not only to ground based astronomical observation.

\section*{Appendix A, Low Earth Orbit Constellations technical details}
This compendium comes from [41] and [42] publications.\\
\begin{figure}[htbp]
\centering
\fbox{\includegraphics[width=0.9\linewidth]{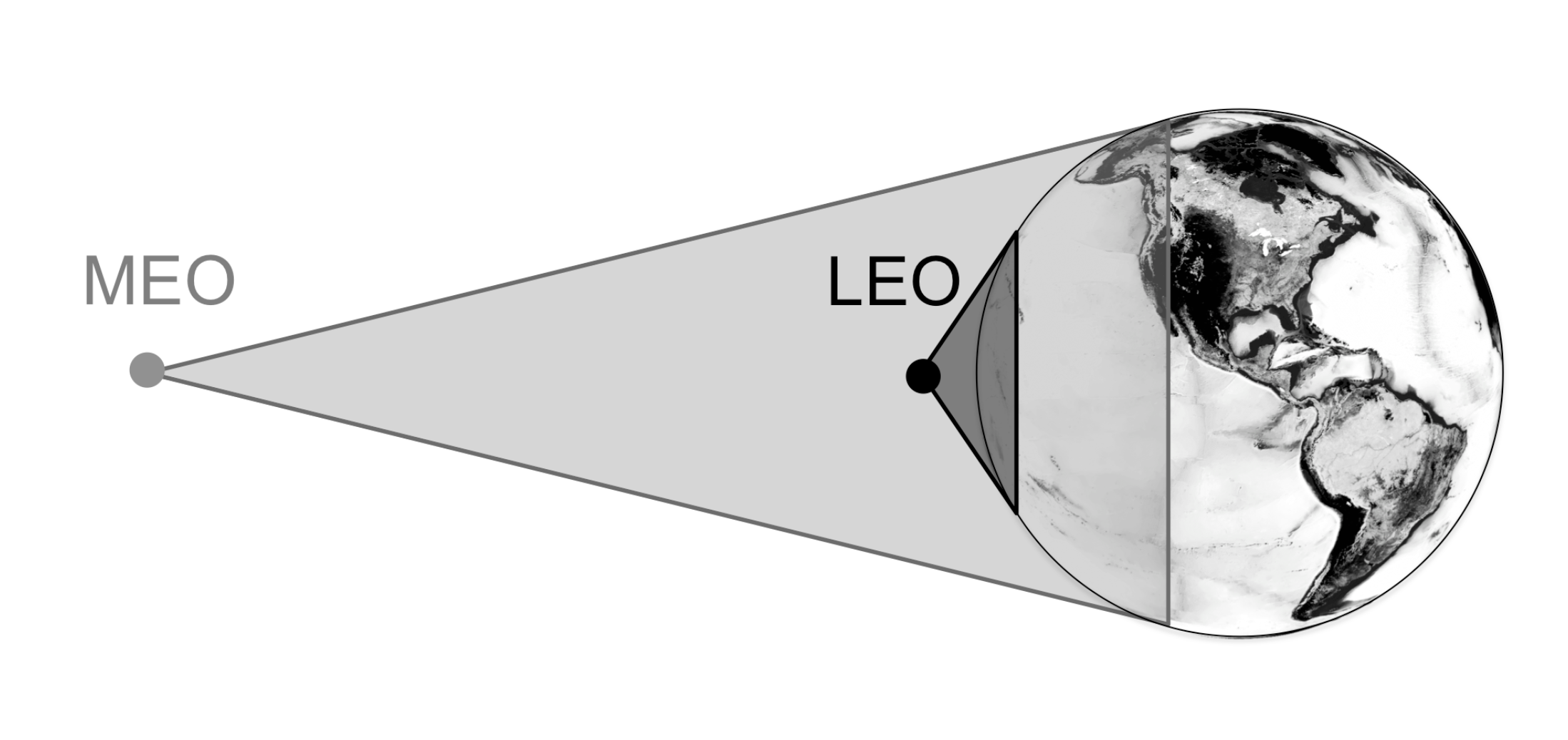}}
\caption{Comparison between a LEO satellite and a MEO satellite in terms of ground coverage.}
\label{fig:leomeo}
\end{figure}
\begin{figure}[htbp]
\centering
\fbox{\includegraphics[width=0.9\linewidth]{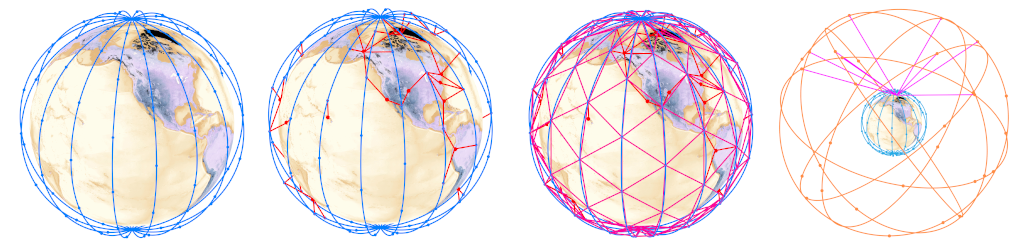}}
\caption{ Each orbit determination puts some constraints on the ground-based stations, and this puts other constraints on the cross-link between satellites to synchronize time and orbits; for this purpose the GPS MEO satellites are used. }
\label{fig:satlinks}
\end{figure}

\begin{figure*}[t]
\centering
\fbox{\includegraphics[width=0.9\linewidth]{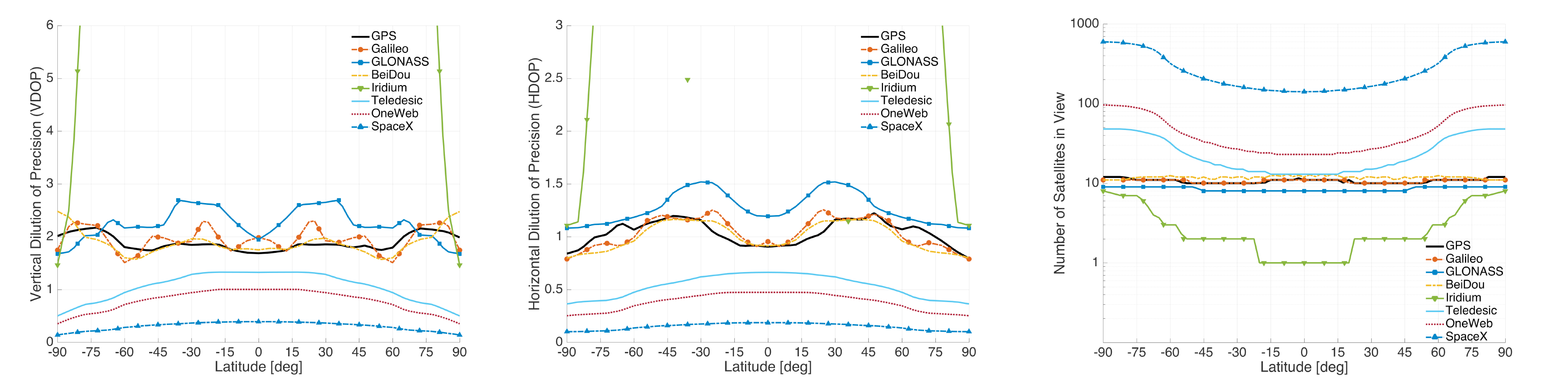}}
\caption{ Vertical (L) and Horizontal (M) Dilution of Precision as a function of latitude for LEO and MEO, and median visibility (R). }
\label{fig:leovisibility}
\end{figure*}

\begin{figure}[htbp]
\centering
\fbox{\includegraphics[width=0.9\linewidth]{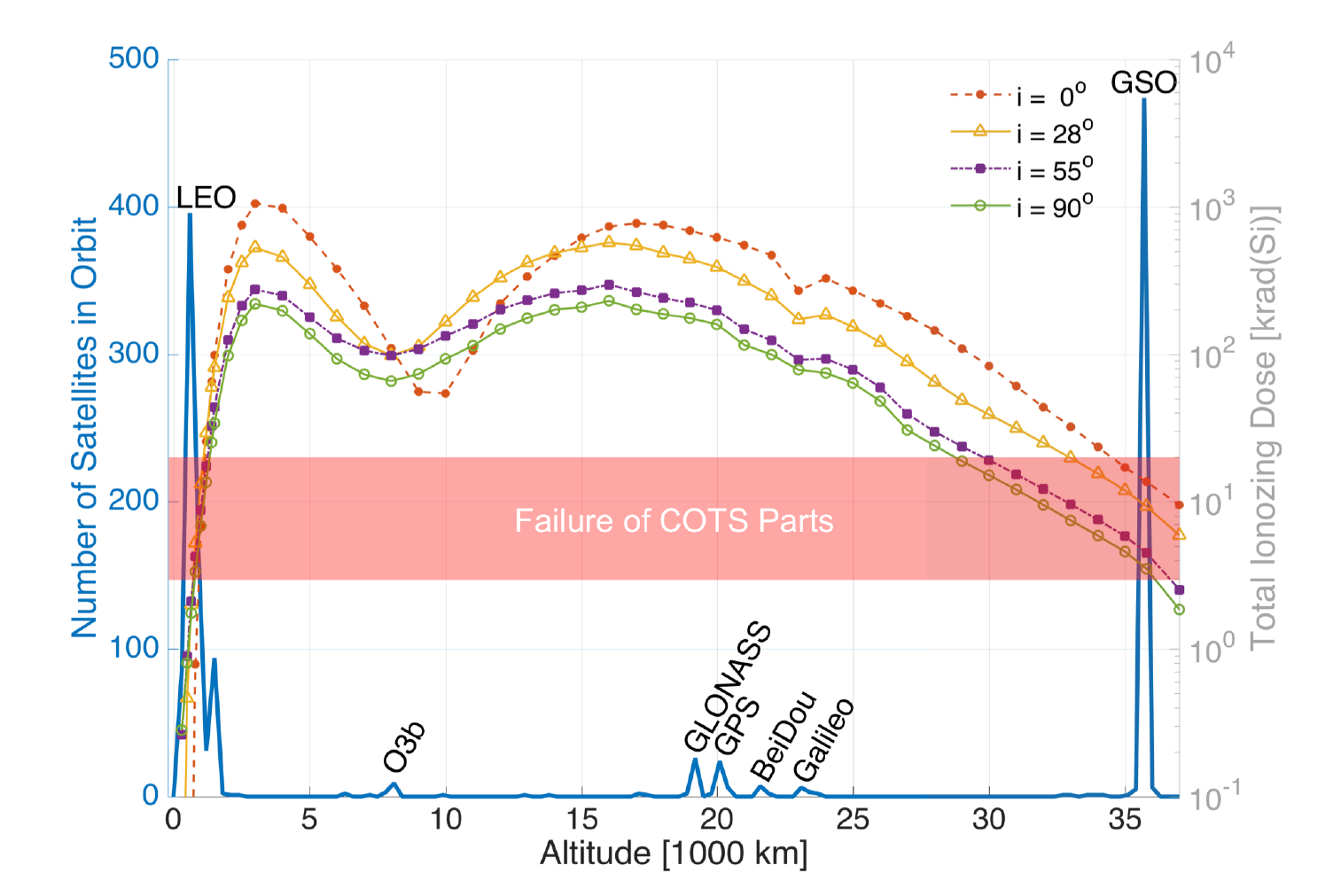}}
\caption{The medium time before degradation of a LEO satellite depends on the altitude; for very big projects like SpaceX it is foreseen to replace 4,000-8,000 Starlink satellites every year after the first five years.}
\label{fig:degradation}
\end{figure}

There are three distinct regimes where satellites tend to reside: 
\begin{enumerate}
\item \textbf{Low Earth Orbiting, LEO} satellites are generally placed between 400 and 1,500 km in altitude; lower orbits decay too quickly due to atmospheric drag, while higher orbit lifespan is cut short from radiation exposure in the lower Van Allen belt.
\item \textbf{Medium Earth Orbiting, MEO} satellites are placed between LEO and GSO, so above 1,500 km in altitude and below 35,000 km. 
\item \textbf{Geo Synchronous Orbiting, GSO} satellites are placed just above 35,000 km; they are designed with orbital periods that matches that of Earth; if placed at the equator, these satellites stay in the same place in the sky.
\end{enumerate}

The total number of satellites for telecommunications are very different depending on the service to provide and the altitude, see Table 2. It is possible to focus the attention only to a few constellation projects.

There is nearly an equal divide of the 1,419 operational satellites in orbit today, between LEO with 780 (+300 Starlink) and GSO with 506. There is not much placed in the vast distance between, nearly three Earths apart. 

Between LEO and GSO, radiation levels are high, so MEO satellites demand specialized hardened components, so the number of these satellites is relatively low (\(\sim100\)) and are dedicated mainly to navigation GPS services.
\\

The Iridium constellation today consists of 66 LEO polar satellites. Fig. 19 shows the Iridium constellation in comparison with other LEO constellations and the current 31 satellite GPS constellation in MEO. It also shows the scale of the difference in altitude with Iridium at 780 km and GPS at 20,200 km.\\
The lower the orbital plane, the more satellites are needed to cover the Earth in LEO than in MEO (Fig. 20). This was one of the fundamental trade-offs considered in the design of the GPS constellation; but also, the higher the altitude, the higher the cost of each launch (Fig. 21).\\
On the other hand, the lower the altitude, the more satellites have to be built to provide coverage. To put this in perspective, global coverage for one satellite in view at all times requires \underline{less than 10 satellites in MEO} but requires \underline{closer to 100 in LEO}.\\

The Broadband LEO constellations have their optimal geometry at the poles. This is because the satellites are in polar orbits, meaning most of the constellation orbits in high latitude regions. This is shown in Fig. 23, which illustrates the typical number of satellites in view as a function of latitude. The Broadband LEOs again perform better in terms of number of satellites visible and their HDOP and VDOP, see Fig. 22.
\begin{figure}[htbp]
\centering
\fbox{\includegraphics[width=0.9\linewidth]{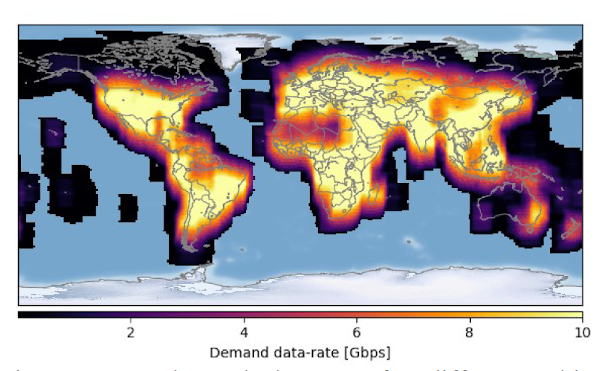}}
\caption{User data demand rate for different orbital positions.}
\label{fig:demandrate}
\end{figure}

Complication in orbital environment reflects in the proper use of cross-link between LEO satellites and with the use of GPS satellites for clock synchronization.\\
Navigation from LEO gives more strength because of the number of elements: more satellites and better geometry allows looser constraints on the orbit and clock.

Depending on the altitude, radiation environment allows for 'careful' cost design.\\
Using closer satellites means stronger signals and resistance to jamming.\\
LEOs satellites move across the sky faster than MEOs, giving some multi-path rejection and covering base-stations better than geo-stationary satellites.\\

Constellation is more robust and fault tolerant than single satellite displacement. This decision was deliberate, as most of the current LEO constellation proposals emphasize offering global bandwidth access for end-users.\\ 
To understand the real demand of the population requesting the service, a demand model needs to be created. A map grid of resolution \( 0.1\, deg -  0.1\, deg \) in latitude and longitude, can be generated in order to determine the number of people covered by the beams of a satellite located in a particular orbital position.

Assuming that any of the satellites will capture at most 10\% of the market at each cell of the grid, a map of the demanded data rate can be created in Fig. 24.\\
\begin{figure}[htbp]
\centering
\fbox{\includegraphics[width=0.9\linewidth]{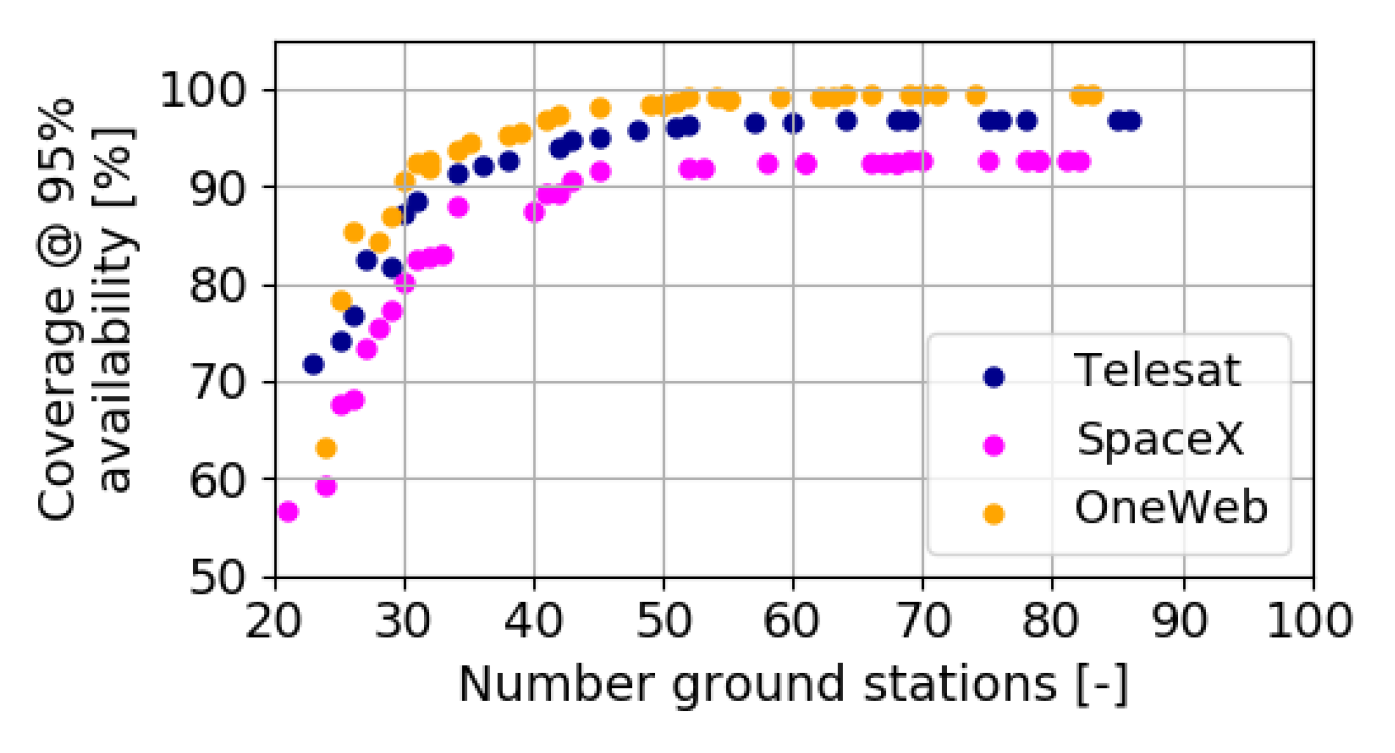}}
\caption{Number of ground station locations vs. demand region coverage.}
\label{fig:groundstations}
\end{figure}
To maximize the throughput of the systems depending on the demand rate and the real capacity of each constellation, it is foreseen to limit the maximum number of ground stations/gateway antennas per site to 50, even though a high degree of coordination among antennas would be required to operate without interference, see Fig. 25.\footnote{Starting from this analysis, the SpaceX company will ask US FCC to upgrade the project with another 30,000 satellites in the V band, enhancing bandwidth and throughput as well. The number of foreseen ground stations is therefore increased to hundreds/thousands for each orbiting satellite.}
\begin{figure*}[t]
\centering
\fbox{\includegraphics[width=0.9\linewidth]{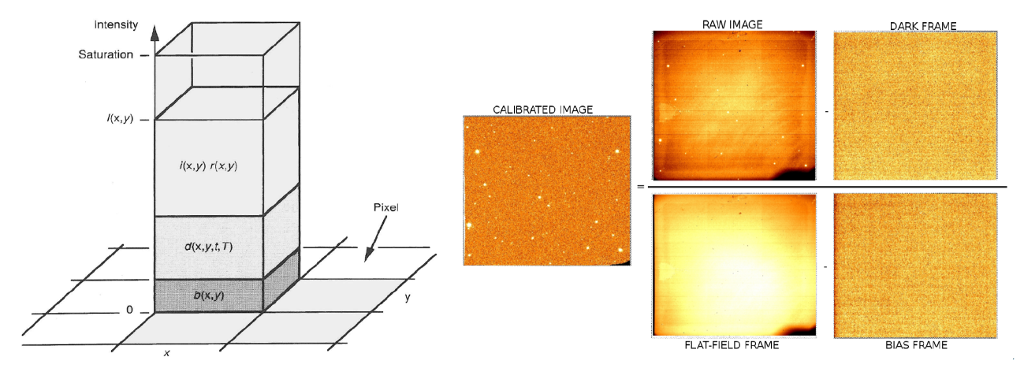}}
\caption{ The principal contributions to (x,y) pixel-counts comes from the nature of CCD output, and can be reduced/calibrated only with good procedures and related calibration frames. }
\label{fig:calibrations}
\end{figure*}

With the full configuration, SpaceX will reach 23.7 Tbps with 123 ground stations (for only 4,425 satellites), thus requiring an extremely large number of ground segments, with hundreds of ground stations and about 3,500 gateway antennas to operate at maximum throughput.\\
Facebook satellites constellation is expected to increase by a factor of 10 the SpaceX performances in terms of data I/O, this can be achieved or using higher frequencies or using more satellites than SpaceX.\\

\section*{Appendix B, Primer of Astronomical Data Reduction}

Considering astronomical observations as a collection of CCD outputs, it is possible to summarise all different intensity contributions as I(x,y), which is measured on the coordinate pixel (x,y) of a RAW image. What is necessary to know is the luminous intensity i(x, y) matrix, that actually illuminates the pixel (x,y).\\ 

The data reduction processing consists in extracting the i(x,y) matrix value from the RAW image matrix, I(x, y).\footnote{When a pair of (or multiple) CCDs are read out simultaneously, one chip (called "killer") affects the counts of other chips ( called "victims"), adding or subtracting ADUs from the real value; so it is necessary to equalize and correct for this effect over all chips of an image and write-out the corrected matrix:\begin{equation}
corrected = victim + xt_{coeff} × (killer_1 + ... + killer_N)
\end{equation}
This is a preliminary reduction step called CrossTalk Correction.}\\
To perform the first step in data-processing, it is possible to identify the three main contributions to the total intensity measured on the RAW image (see Fig. 26):\\
\begin{enumerate}
\item \( b(x,y)\), the BIAS contribution that is the pre-charged value of the CCD; this contribution is
constant and independent of the temperature and the exposure time. It is possible to estimate this contribution by acquiring an exposure in total darkness (with shutter closed) with the shortest
exposure time possible: \begin{equation} lb(x, y) = b(x, y)\end{equation}
\item  \(d(x,y,t,T)\), the DARK contribution that is noise accumulated by thermal loads during the exposure. This contribution is proportional to exposure time \(t\), and for a given exposure time it is smaller as the temperature \(T\) is lowered; it is possible to estimate this contribution by acquiring in total darkness an exposure with the same exposure time and the same temperature of the science image. This kind of exposure contains the bias pre-charged level: 
\begin{equation} 
ld(x,y) = b(x,y) + d(x,y,t,T)
\end{equation}

\item  \(f(x,y)\), the FLAT (RESPONSE) factor of the pixels is only dependent on the (x,y) positions and is estimated by the flat-fields images. To estimate this contribution it is necessary to acquire a set of images of a particular exposure time on a flat illuminated screen/surface, so that \(i(x,y)\) is constant for all flat image pixels. This image contains both the bias and the dark contributions, but the dark may be negligible due to short exposure time of such frames:
\begin{equation*} 
lf (x,y) = b(x,y) + d(x,y,t,T) + f(x,y) · N_{FAC} 
\end{equation*} 
\begin{equation*} 
\rightarrow lf (x,y) = b(x,y) + f(x,y) · N_{FAC} 
\end{equation*} 

The relation that links together the three components is:
\begin{equation}
I(x,y) = b(x,y) + d(x,y,t,T) + i(x,y) · f(x,y)
\end{equation}
The goal of a cosmetic reduction is the extraction of the real signal reaching the CCD, and to do this it is necessary to extract the \(i(x,y)\) contribution:

\begin{equation*} 
i(x,y) = \frac{I(x,y) - [b(x,y) + d(x,y,t,T)]} {f(x, y)} 
\end{equation*}

\begin{equation} 
\rightarrow i(x,y) = { \frac{l(x,y) - ld(x,y)} {ld(x, y) - lb(x, y) } } · N_{FAC}
\end{equation}
\\
\end{enumerate}
Linking the \(b(x,y)\), \(d(x,y,t,T)\) and \(f(x,y)\) to the real calibration images (bias, darks and flat-fields), it is possible to extract the exact value of \(i(x,y)\) matrix in the pre-processing phase, also known as pre-reduction and/or photometric correction.\\

All other data processing, e.g. pixel cosmic rays masking, background subtraction, final astrometric mosaic stack + RMS and exposure map, relies on a good pre-reduction procedure/frames, which depends on the quality of calibration frames used for the whole pipelines chain, see Algorithm 1. \\

Satellites' Constellations pollution can alter the most important data reduction steps (both for images and spectra), compromising the whole scientific content of astronomical ground-based observations. \\

The use of image-masks is fundamental to scientific optical data reduction, because it is possible to avoid spurious counts content in an astronomical image. In particular it has been remarked that it is not possible to replace the altered value with the local background around bad-pixels (trails, or cosmic rays, or ghosts, etc), because the the photometry would be altered too.\\
To preserve photometry it is necessary to produce a parallel masks of "good" and "bad" pixels and these masks have to be updated as each reduction step is performed (e.g. if a cosmic ray falls on some pixels and saturate them, those pixels are flagged to 1, then a "dilate" operation is performed to increase the isoarea of related mask by a factor to be choose (2-4-10): the more is the dilate factor the more is preserved the photometry.\\
In particular trails in fig.4 are related to dilated masks left by the passage of 3rd Mag satellites; trails used to resample and coadd the final mosaic are best seen in the single exposure in Fig. 6.\\
 
We summarize the whole standard "data reduction chain" for LBC data, see Algorithm 1 and [43] for details.\\

\begin{algorithm}[H]
\caption{Standard Scientific Data Reduction Pipeline for LBC: to obtain a photometric calibrated mosaic and RMS map starting from raw ccd images, i(x,y)}
\includegraphics[width=1.0\linewidth]{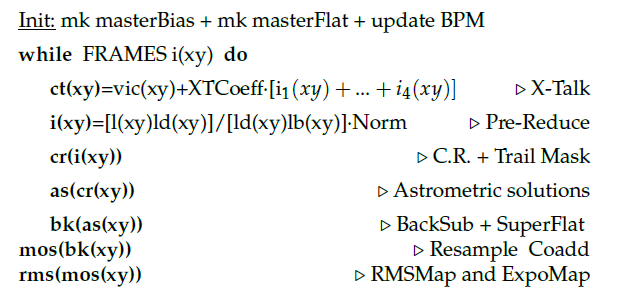}
\end{algorithm}

\section*{Appendix C, Reference Systems Adopted in this work: the LBC case study}

\begin{figure}[htbp]
\centering
\fbox{\includegraphics[width=0.9\linewidth]{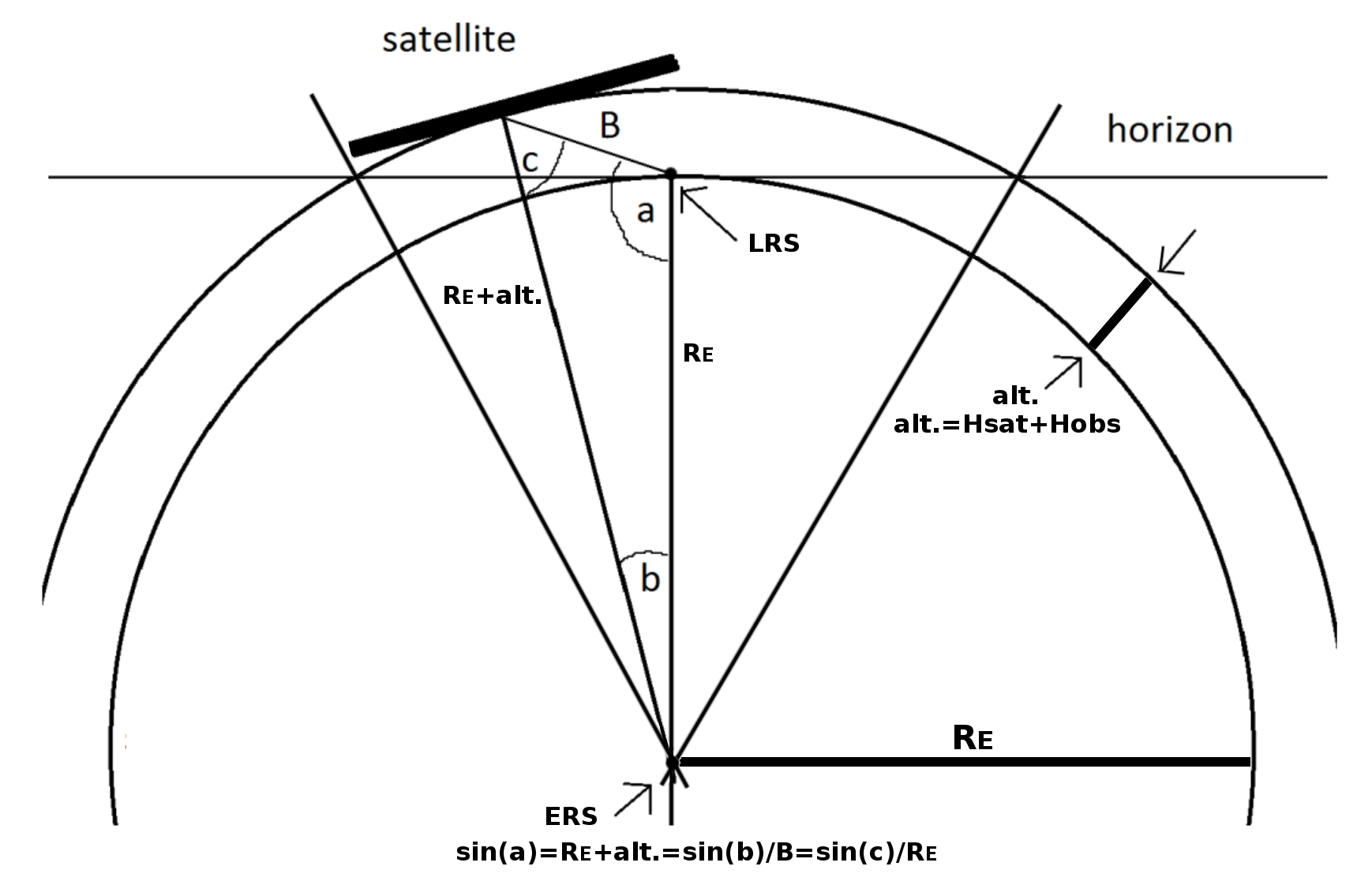}}
\caption{Graphical representation of Reference systems used in this work, see [60].}
\label{fig:refsys}
\end{figure}
The number of satellites passing in an LBC-like field of \(23\,arcmin\,\cdot\,25\,arcmin=0.16\,deg^2\) is the (Number of satellites in the FOV at time t= 0) + (Number of satellites passing in the diagonal of the rectangular FOV during the Total Exposure Time of the Mosaic); for the LBC FOV the diagonal linear dimension is \(\sqrt{23^2+25^2}/60=33/60=0.566\,deg\).\\

So in \(\sim1\,h\) of Total Exposure Time computed in \textbf{ERS}, where \(\rho\simeq 1.2\) is the absolute satellite density:\\
\(N_{FOV}\,(t = 0)\, =\, 0.16 * 1.2\, =\, 0.192\, sats\, at\, t = 0\)\\
\(+\: N_{cross\,in\,1h\,obs} \rightarrow\)\\
In one hour a satellite travels in the sky (in ERS at about 340km altitude) \(= 0.06 * 3600 = 216\, degrees\), so 216 degrees with a density of 1.2 sat per deg \(= 260 sats\, in\, a\, deg^2\), then there will transit within the LBC diagonal \(of\, \sim 0.566\,deg\) \(\sim\,122\, sats\,in\,1h\) or just 1 satellite every \(\sim\,30\, seconds\).\\

The same quantities can be computed in \textbf{LRS} where \(\omega\) is greater by a factor of about 20, but at the same time \(\rho \simeq 0.06\) is the lower local satellite density.\\
\(N_{FOV}\,(t = 0)\, =\, 0.16 * 0.06\, =\, 0.0096\, sats\, at\, t = 0\)\\
\(+\: N_{cross\,in\,1h\,obs} \rightarrow\)\\
In one hour a satellite travels in the sky (in LRS at about 340km altitude) \(= 1.32 * 3600 = 4752\, degrees\), so 4752 degrees with a density of 0.06 sat per deg \(= 285 sats\, in\, a\, deg^2\), then there will transit within the LBC diagonal \(of\,\sim 0.566\, deg\)  \(\sim\,161\, sats\,in\,1h\) or just 1 satellite every 22 seconds.\\

\underline{Numbers are comparable} (indeed in LRS trails are slightly more frequent) that's why in a conservative way authors decided to keep the ERS by changing the observation point of view and greatly simplifying all the stuff.\\

\section*{ APPENDIX D, International Conventions and Treaties: a legal approach }

The Preamble of the \textbf{World Heritage Convention} holds that
\textquotedblleft the deterioration or disappearance of any item of
the cultural or natural heritage constitutes a harmful impoverishment
of the heritage of all the nations of the world\textquotedblright{}
This protection appears again in the 1994 Universal Declaration of
Human Rights for Future Generations:
\begin{quote}
Persons belonging to future generations have the right to an
uncontaminated and undamaged Earth, including pure skies; they are
entitled to its enjoyment as the ground of human history of culture
and social bonds that make each generation and individual a member
of one human family.
\end{quote}

The \textbf{UNESCO} has undertaken activities for the safeguarding
of cultural heritage related to astronomy under the \textquotedblleft Astronomy
and World Heritage\textquotedblright{} project launched by the World
Heritage Centre in 2003. This concept was taken up again by UNESCO
in 2005 as:
\begin{quote}
The sky, our common and universal heritage, is an integral part
of the environment perceived by humanity. Humankind has always observed
the sky either to interpret it or to understand the physical laws
that govern the universe. This interest in astronomy has had profound
implications for science, philosophy, religion, culture and our general
conception of the universe.
\end{quote}

This in turn led to the following concepts:
\begin{quote}
Astronomical observations have profound implications for the
development of science, philosophy, religion, culture and the general
conception of the universe\dots{} discoveries of astronomers in the
field of science have had an influence not only on our understanding
of the universe but also on technology, mathematics, physics and social
development in general\dots{} the cultural impact of astronomy has
been marginalized and confined to a specialized public.
\end{quote}

These protections for Starlight are necessary as the impact that Starlight
has held on humanity has been expressed in works of religion, art,
literature, science, philosophy, business, and travel. To note the further enforcement of the \underline{Right to Starlight}:
\begin{quote}
International law enforces international legal obligations,
including property interests. Here, World Heritage is the property
of all humankind, and while there may be protective laws, enforcing
this is another matter, as only States can sue other States under
this type of international treaty.
\end{quote}

\textit{So a State is responsible for the activities that occur within its jurisdiction
-- whether they are authorized or unauthorized.}
\begin{quote}
    Within the framework of International Law and State based legal
instruments, Protection of Starlight could then be implemented in
the same manner:\\
1. \textit{Reaffirms the sovereign rights and responsibilities, towards
the International Community, of each State for the protection of its
own cultural and natural heritage;}\\
2. \textit{Calls upon the International Community to provide all the
possible assistance needed to protect and conserve the cultural and
natural heritage of Starlight;}\\
3. \textit{Invites the authorities of States to take appropriate measures
in order to safeguard the cultural and natural heritage of Starlight;}\\
4. \textit{Further invites the States to co-operate with UNESCO, the
World Heritage Committee, the UNWTO, and the Starlight Initiative
with a view to ensuring effective protection of its cultural and natural
heritage in Starlight.}
\end{quote}

Having established these rights under international law, the conclusion
is that there exist duties for both States and international organizations
to \underline{protect the World Heritage Right to Starlight}, as well as, their
duties to foster the rights of travelers, hosts, and providers of
travel to enjoy this \textit{Starlight "property interest" that belongs to all humanity}. \\
The existing legal instruments demonstrate the protection for the Right to Starlight, but it is the States that act as custodians of World Heritage that are charged with ensuring
these rights are enforceable, and in turn made available to all of humanity.

\subsection*{Legal Considerations}

SpaceX and other private company received permission from any government
agencies (e.g. Federal Communication Commission, FCC for USA) to launch these
satellites into orbit. So there could be a legal claim, within the
US legal system, to halt the progress of Starlink fleet.\\

Also, as it turns out, according to the Outer Space Treaty and its
progeny, there are no private companies operating in outer space,
but only governments can operate in outer space. And the legal process
is that the state government, this time the USA government, is legally
responsible for all objects sent into outer space that launch from
USA borders. That means, that it is the USA government that is responsible
for the harm caused by its corporation, Starlink, sending objects
into orbit that cause harm. \\

So under this international law, any country that suffers harm by
Starlink can sue the United States (or any other) government in the International
Court of Justice in the Hague. The harm here is damage to our cultural
heritage, the night sky, and monetary damages due to the loss of radio
and other types of astronomy. For the scientists, the owners of the
observatories have a legal argument that they have and will continue
to lose money spent for their research based on Earth based observatories.
Furthermore, Universities that own the observatories are state owned
universities, so it is the government that owns the observatories
that have lost financially because of their interruption of study
of the night skies.\\

So it is essential that a government, like Chile, Italy or France,
sues the USA in the International Court of Justice to halt deployment of satellites' mega-constelations.\\

\textbf{If no national or international entity will stop this displacement,
the right of the private companies (e.g. SpaceX) will become
acquired at the beginning of March/April 2020}.\\

How should the international astronomical community mobilize in order
to stop further Starlink launches?
\begin{enumerate}
\item Sue in court for luminous pollution not taken into account by US FCC:
The FCC's lack of review of these commercial satellite
projects violates the National Environmental Policy Act, NEPA, which
obligates all federal agencies to consider the environmental impacts
of any projects they approve. So in the most basic sense, SpaceX's
satellites displacement authorization would be unlawful, see {[}31{]}.
\item Sue in court for lack of jurisdiction and jurisprudence of US FCC
to authorize private not geostationary satellites over other states
and nations.
\item Sue in the International Court of Justice, ICJ the USA government
to put on hold further Starlink launches to quantify the loss of public
finances in damaging national and international astronomical projects.\footnote{Though there are no international law that restrict mega constellations,
to deploy and dispatch mega constellations an international agreement
among states is needed, since satellites can not be located only over
a single state (e.g. USA) but, being in LEO, they move around the
globe passing over different states/nations/continents. This is a
lack of jurisdiction of FCC authorization. In particular the International
Court of Justice, ICJ, can be called into question whenever there
is a dispute of international jurisdiction or between member states
of the United Nations on the basis of international norms, treaties
and / or their violations. In the beginning of chapter 5 it was explained
how the World Heritage Convention regarding the ``right of night
sky / starlight'' belongs to universal human rights and so no state
can decide to contravene this convention if it interferes with the
enjoyment of that right for other states. The pretext for appealing
to the United Nations and the International Court of Justice (ICJ)
is the loss of scientific value of the investments made for ground
based projects by each state (damaged by SpaceX). Each damaged state,
being damaged as conseguence of a violation for an international treaty,
the issue cannot be settled with a simple money compensation, but
with an inhibition of the damage before the same occurs (and not after).} 
\end{enumerate}

\subsection*{What can be done by Astronomers}

An \textbf{international appeal/petition from astronomers} was launched
in January 2020 and, at the time of writing, thousands of astronomers
involved with astronomical observatories and facilities, have subscribed
the appeal, see {[}29{]}. Another \textbf{open letter} has been prepared
reguarding same concerns on the satellites constellations deployment
for the further space missions and to raise awareness to US Senate,
and US commisisons on the possibility that occurs the Kessler Syndrome, see {[}34{]},
which is a realistic scenario with all these orbiting objects, see
{[}36{]}.\\ 

Requests from the astronomical community to governments, institutions,
and agencies all around the world are:
\begin{enumerate}
\item to be committed to provide legal protection to ground astronomical
facilities in all of the available observation electromagnetic windows. 
\item to put on hold further Starlink launches (and other projects) and
carry out an accurate moratorium on all technologies that can negatively
impact astronomical space based and ground based observations, or
impact on the scientific, technological and economic investments that
each State engages in astrophysical projects. 
\item to put in place a clear evaluation of risks and predictive impacts
on astronomical observatories (i.e. loss of scientific and economic
value), giving stringent guidelines to private individuals, societies
and industries to plan satellite investments without clearly understanding
all of the negative effects on outstanding astronomical facilities. 
\item that the US Federal Communications Commission (FCC) and any other
national agency be wary of granting permission to ship non-geostationary
low-orbit satellites into orbit or alternatively to limit the authorization
of only satellites being above the airspace of the \textquotedblleft home
country\textquotedblright . 
\item to demand a worldwide orchestration, where national and international
astronomical agencies can impose the right of veto on all those projects
that negatively interfere with astronomical outstanding facilities. 
\item to limit and regulate the number of telecommunication satellite fleets
to the \textquotedblleft strictly necessary number\textquotedblright{}
and to put them in orbit only when old-outdated technology satellites
are deorbited, according to the Outer Space Treaty (1967) - the Art
IX, and the United Nations Guidelines for the Long-term Sustainability
of Outer Space Activities (2018) -- guideline 2.2(c), requiring
the use of outer space be conducted \textquotedblleft so as to avoid
{[}its{]} harmful contamination and also adverse changes in the environment
of the Earth\textquotedblright{} and {[}\dots omissis\dots{]} risks
to people, property, public health and the environment associated
with the launch, in-orbit operation and re-entry of space objects\textquotedblright .
\end{enumerate}

\section*{Acknowledgments}The authors wish to thank all INAF colleagues for their effort, recommendations and technical support about different topics of this work.\\
A special thanks is dedicated to Ben Levi for the editorial effort in the second version of this preprint. 

\section*{Notes} During the submission of this preprint at the same time a new article by ESO (Olivier R. Hainaut and Andrew P. Williams) was submitted to A\&A, with different methodologies and different conclusions. It is not simple to compare such a different approach, but it is necessary to remark that in the ESO paper there is a huge underestimation in the total number of planned satellites, and neither VISTA nor VST were considered, while too much emphasis is given to telescopes with very short FOV, like VLT and ELT, where it is clear that statistically few satellites will fall into them. \\At the following link it is possible to find the ESO article:  \href{https://arxiv.org/abs/2003.01992}{arxiv.org/abs/2003.01992}.  
Another article by CFA (Jonathan C. McDowell) was published specially related to Starlink Fleet; conclusions are similar to ESO's paper, unfortunately it is necessary to note that only 12,000 starlink sats were considered, so the same concerns of ESO's paper have to ve arisen. \\At the following link it is possible to find the CFA article:  \href{https://arxiv.org/abs/2003.07446}{arxiv.org/abs/2003.07446}. 

\section*{References}

\begin{flushleft}
\texttt{{[}1{]} S.Gallozzi, M.Scardia, M.Maris, "concerns about ground-based astronomical observations: a step to safeguard the astronomical sky", feb.2020, arXiv:2001.10952:  \href{https://arxiv.org/pdf/2001.10952.pdf}{https://arxiv.org/pdf/2001.10952.pdf}} 
\par\end{flushleft}

\begin{flushleft}
\texttt{{[}2{]} International Astronomical Union, IAU 1st statements: \href{https://www.iau.org/news/announcements/detail/ann19035/?lang}{www.iau.org/ann19035}} 
\par\end{flushleft}

\begin{flushleft}
\texttt{{[}3{]} International Astronomical Union, IAU 2nd statements: \href{https://www.iau.org/news/pressreleases/detail/iau2001/}{www.iau.org/iau2001}} 
\par\end{flushleft}

\begin{flushleft}
\texttt{{[}4{]} ESO Very Large Telescope: \href{https://www.eso.org/public/usa/teles-instr/paranal-observatory/vlt/?lang}{www.eso.org/paranal-observatory/vlt}}
\par\end{flushleft}

\begin{flushleft}
\texttt{{[}5{]} Large Binocular Telescope, LBT: \href{http://www.lbto.org/}{http://www.lbto.org/}}
\par\end{flushleft}

\begin{flushleft}
\texttt{{[}6{]} Large Binocular Camera, LBC: \href{http://lbc.oa-roma.inaf.it/}{http://lbc.oa-roma.inaf.it/}}
\par\end{flushleft}

\begin{flushleft}
\texttt{{[}7{]} ESO Extremely Large Telescope, E-ELT: \href{https://www.eso.org/public/usa/teles-instr/elt/?lang}{www.eso.org/elt} and Very Large Telescope, VLT: \href{https://www.eso.org/public/usa/teles-instr/paranal-observatory/vlt/?lang}{www.eso.org/vlt}}
\par\end{flushleft}

\begin{flushleft}
\texttt{{[}8{]} Large Syhoptic Survey Telescop, LSST: \href{https://www.lsst.org\%20\%E2\%80\%93\%20https://en.wikipedia.org/wiki/Vera_C._Rubin_Observatory}{Vera\_C.\_Rubin\_Observatory}}
\par\end{flushleft}

\begin{flushleft}
\texttt{{[}9{]} ESO VLT Survey Telescope: \href{https://www.eso.org/public/italy/teles-instr/paranal-observatory/surveytelescopes/vst/}{surveytelescopes/vst}}
\par\end{flushleft}

\begin{flushleft}
\texttt{{[}10{]} Pan-STARRS Telescope: \href{https://panstarrs.stsci.edu/}{https://panstarrs.stsci.edu}}
\par\end{flushleft}

\begin{flushleft}
\texttt{{[}11{]} Keck Observatory: \href{http://www.keckobservatory.org/}{http://www.keckobservatory.org}}
\par\end{flushleft}

\begin{flushleft}
\texttt{{[}12{]} United Nations Office for Outer Space Affairs, ``Treaty
on Principles Governing the Activities of States in the Exploration
and Use of Outer Space, including the Moon and Other Celestial Bodies'': \href{https://www.unoosa.org/oosa/en/ourwork/spacelaw/treaties/introouterspacetreaty.html}{introouterspacetreaty.html}}
\par\end{flushleft}

\begin{flushleft}
\texttt{{[}13{]} Committee on the Peaceful Uses of Outer Space, ``Guidelines
for the Long-term Sustainability of Outer Space Activities '': \href{https://www.unoosa.org/res/oosadoc/data/documents/2018/aac_1052018crp/aac_1052018crp_20_0_html/AC105_2018_CRP20E.pdf}{https://www.unoosa.org}}
\par\end{flushleft}

\begin{flushleft}
\texttt{{[}14{]} Simulated prediction of \textquotedblleft only\textquotedblright{}
12k Starlink satellites positions in the sky: \href{https://youtu.be/LGBuk2BTvJE}{www.youtu.be/LGBuk2BTvJE} }
\par\end{flushleft}

\begin{flushleft}
\texttt{{[}15{]} Simulated prediction of \textquotedblleft only\textquotedblright{}
12k Starlink satellites naked eye view of the sky: \href{https://www.youtube.com/watch?v=z9hQfKd9kfA}{www.youtube.com/9hQfKd9kfA}}
\par\end{flushleft}

\begin{flushleft}
\texttt{{[}16{]} Visualization tool to find, plot and search satellite
orbits: \href{https://celestrak.com/cesium/orbit-viz.php}{https://celestrak.com/orbit-viz}}
\par\end{flushleft}

\begin{flushleft}
\texttt{{[}17{]} Phil Cameron, ``The right to Starlight Under International Law''}
\par\end{flushleft}

\begin{flushleft}
\texttt{{[}18{]} Starlight Initiative ``Declaration in defense of the night sky and the right to starlight'': \href{https://issuu.com/pubcipriano/docs/starlightdeclarationen}{starlightdeclaration}}
\par\end{flushleft}

\begin{flushleft}
\texttt{{[}19{]} Marian Cipriano, ``Starlight: a common heritage'', Published online by Cambridge University Press: 29 June 2011 }
\par\end{flushleft}

\begin{flushleft}
\texttt{{[}20{]} Marin, C., and Jafar, J. (eds) 2008,Starlight: A common Heritage(Tenerife: Instituto de As-trof\'{ }\i sica de Canarias)}
\par\end{flushleft}

\begin{flushleft}
\texttt{{[}21{]} Starlight Scientific Committee Report 2009,Starlight Reserve Concept: \href{http://www.starlight2007.net/pdf/StarlightReserve.pdf}{http://www.starlight2007.net}}
\par\end{flushleft}

\begin{flushleft}
\texttt{{[}22{]} A.J. Beasley, E. Murphy, R. Selina, M. McKinnon \& the ngVLA Project Team, ``The Next Generation Very Large Array (ngVLA)'' NRAO}
\par\end{flushleft}

\begin{flushleft}
\texttt{{[}23{]} Square Kilometer Array, SKA: \href{https://www.skatelescope.org/}{https://www.skatelescope.org}}
\par\end{flushleft}

\begin{flushleft}
\texttt{{[}24{]} Avinash A. Deshpande and B. M. Lewis, ``Iridium Satellite Signals: A Case Study in Interference Characterization and Mitigation for Radio Astronomy Observations'', DOI: 10.1142/S225117171940009, arXiv:1904.00502 {[}astro-ph.IM{]}}
\par\end{flushleft}

\begin{flushleft}
\texttt{{[}25{]} Committee on Radio Frequencies Board on Physics and Astronomy Commission on Physical Sciences, Mathematics, and Applications National Research Council ``VIEWS OF THE COMMITTEE ON RADIO FREQUENCIES CONCERNING FREQUENCY ALLOCATIONS FOR THE PASSIVE SERVICES AT THE 1992 WORLD ADMINISTRATIVE RADIO CONFERENCE `` }
\par\end{flushleft}

\begin{flushleft}
\texttt{{[}26{]} Atacama Large Millimeter Array, ALMA: \href{https://www.eso.org/public/usa/teles-instr/alma/?lang}{https://www.eso.org}}
\par\end{flushleft}

\begin{flushleft}
\texttt{{[}27{]} SKA statements on satellites contellations: \href{https://www.skatelescope.org/news/ska-statement-on-satellite-constellations/}{ska-statement-on-satellite-constellations}}
\par\end{flushleft}

\begin{flushleft}
\texttt{{[}28{]} PATRICIA COOPER, ``SATELLITE GOVERNMENT AFFAIRS SPACE EXPLORATION TECHNOLOGIES CORP. (SPACEX)'': \href{https://www.commerce.senate.gov/services/files/6c08b6c2-fe74-4500-ae1d-a801f53fd279}{https://www.commerce.senate.gov}}
\par\end{flushleft}

{[}29{]} APPEAL BY ASTRONOMERS: \href{https://astronomersappeal.wordpress.com}{https://astronomersappeal.wordpress.com}\\

{[}30{]} Starlink testing encrypted internet form military US Air
Force purposes: \href{https://www.reuters.com/article/us-spacex-starlink-airforce/musks-satellite-project-testing-encrypted-internet-with-military-planes-idUSKBN1X12KM}{https://www.reuters.com/article}\\

{[}31{]} Starlink APPLICATION FOR APPROVAL FOR ORBITAL DEPLOYMENT AND OPERATING AUTHORITY FOR THE SPACEX NGSO SATELLITE SYSTEM: \href{https://cdn.arstechnica.net/wp-content/uploads/2017/05/Legal-Narrative.pdf\%20}{Legal-Narrative.pdf }\\

{[}32{]} Starlink simulations from deepskywatch: \href{http://www.deepskywatch.com/Articles/Starlink-sky-simulation.html}{http://www.deepskywatch.com}\\

{[}33{]} Volvach et al 2019, ``An Unusually Powerful Water-Maser Flare in the Galactic Source W49N'', doi: \href{https://ui.adsabs.harvard.edu/link_gateway/2019ARep...63..652V/doi:10.1134/S1063772919080067}{2019ARep...63..652V/doi:10.1134}

{[}34{]} Donald J. Kessler, Burton G. Cour-Palais, ``Collision frequency of artificial satellites: The creation of a debris belt'', doi: \href{https://doi.org/10.1029/JA083iA06p02637}{https://doi.org/10.1029/JA083iA06p02637} and direct link: \href{http://www.castor2.ca/07_News/headline_010216_files/Kessler_Collision_Frequency_1978.pdf}{.pdf}

{[}35{]} Ramon J. Ryan, Note, The Fault In Our Stars: Challenging the FCC\textquoteright s Treatment of Commercial Satellites as Categorically Excluded From Review Under the National Environmental Policy Act,
22 VAND. J. ENT. \& TECH. L. (forthcoming May 2020)

{[}36{]} David Dubois, NASA ``Open Letter to FCC, US Senate and Commissions and SpeceX``, \href{https://docs.google.com/document/d/1fScGypCWEG6j2S7oPBb8u0ibQ5lRAXwkG4Z4zj2pGig/edit}{Open LEtter for FCC and NASA} 

{[}37{]} Diego Paris, S. Gallozzi, V. Testa ``The processing system for the reduction of the INAF LBT imaging data``, \href{http://abell.as.arizona.edu/~lbtsci/UM2017/Presentations/LBTO_meeting_Diego_Paris.pdf}{Presentation to LBTO 2017 Users' Meeting, Florence, 20-23 June 2017} 

{[}38{]}  Seitzer, Pat (University of Michigan), 2020, "Presentation to the US National Science Foundation Astronomy and Astrophysics Advisory Committee", \href{https://www.nsf.gov/attachments/299316/public/12_Satellite_Constellations_and_Astronomy-Pat_Seitzer.pdf}{Presentation to US NSFAAA Committee} 

{[}39{]}  Inigo del Portillo,  Bruce G. Cameron, Edward F. Crawley, "A Technical Comparison of Three Low Earth Orbit Satellite Constellation Systems to Provide Global Broadband", 69th International Astronautical Congress 2018, "A Technical Comparison of Three Low Earth Orbit Satellite Constellation Systems to Provide Global Broadband", \href{http://www.mit.edu/~portillo/files/Comparison-LEO-IAC-2018-slides.pdf}{www.mit.edu/Comparison-LEO-IAC-2018-slides} 

{[}40{]}  Sardinia Radio Telescope , SRT: \href{http://www.srt.inaf.it/}{http://www.srt.inaf.it} 

{[}41{]} Inigo del Portillo, Bruce G. Cameron et al, MIT, 2018 "A Technical Comparison of Three Low Earth Orbit Satellite Constellation Systems to provide Internet Broadband" \href{https://doi.org/10.1016/j.actaastro.2019.03.040}{doi.org/10.1016/j.actaastro.2019.03.040} 

{[}42{]} Tyler G. R. Reid, Andrew M. Neish, Todd F. Walter, \& Per K. Enge
Stanford University, MIT, 2018 "Leveraging Commercial Broadband LEO Constellations for Navigation" \href{https://doi.org/10.1002/navi.234}{doi.org/10.1002/navi.234} and presentations: \href{http://web.stanford.edu/group/scpnt/pnt/PNT16/2016_Presentation_Files/S11-Reid.pdf}{presentation}.

{[}43{]} Stefano Gallozzi, INAF-Osservatorio Astronomico di Roma, 2012 "Physical Data Reduction Methods" \href{https://drive.google.com/file/d/1BrXklBKRTXwR7ch5IWr5HXYrHW6ykMby/view?usp=sharing}{googledoc}.

{[}44{]} INAF, "Piano della Performance dell’INAF (2017 –2019)": \href{https://performance.gov.it/performance/piani-performance/documento/788}{https://performance.gov.it}.

{[}45{]} MeerKAT, South Africa Radio Telescope: \href{https://www.sarao.ac.za/science-engineering/meerkat/}{https://www.sarao.ac.za/science-engineering/meerkat/}.  

{[}46{]} ASKAP, Australian Square Kilometer Array Pathfinder: \href{https://www.atnf.csiro.au/projects/askap/science.html}{www.atnf.csiro.au/projects/askap}.  
{[}47{]} ASTRI and ASTRI MiniArray: \href{http://www.brera.inaf.it/~astri/wordpress/}{www.brera.inaf.it/~astri}. 

{[}48{]} MAGIC TELESCOPES: \href{https://magic.mpp.mpg.de}{magic.mpp.mpg.de}. 

{[}49{]} CTA Observatory: \href{https://www.cta-observatory.org/}{www.cta-observatory.org}. 

{[}50{]} H.E.S.S Telescope: \href{https://www.mpi-hd.mpg.de/hfm/HESS/}{www.mpi-hd.mpg.de/hfm/HESS}. 

{[}51{]} VERITAS Array: \href{https://veritas.sao.arizona.edu/}{veritas.sao.arizona.edu}. 

{[}52{]} UNITED NATIONS OFFICE FOR OUTER SPACE AFFAIRS, "Space Debris Mitigation Guidelines of the Committee  on the Peaceful Uses of Outer Space": \href{"http://www.unoosa.org/pdf/publications/st_space_49E.pdf"}{www.unoosa.org/pdf}. 

{[}53{]} The Guardian, "We've left junk everywhere: why space pollution could be humanity's next big problem": \href{https://www.theguardian.com/science/2017/mar/26/weve-left-junk-everywhere-why-space-pollution-could-be-humanitys-next-big-problem}{the guardian link}. 

{[}54{]} The Guardian, "Junk from China missile test raises fear of satellite collision": \href{https://www.theguardian.com/science/2007/feb/07/spaceexploration.weekendmagazinespacesection}{the guardian link}. 

{[}55{]} New Scientist, "India tests anti-satellite missile by destroying one of its satellites": \href{https://www.newscientist.com/article/2197903-india-tests-anti-satellite-missile-by-destroying-one-of-its-satellites/}{New Scientist link}. 

{[}56{]} International Dark Sky Association, "IDA Responds to Satellite Megaconstellations": \href{https://www.darksky.org/ida-responds-to-satellite-megaconstellations/}{IDA statement link}. 

{[}57{]} Forbes: \href{https://www.forbes.com/sites/startswithabang/2019/11/20/this-is-how-elon-musk-can-fix-the-damage-his-starlink-satellites-are-causing-to-astronomy/#52261a044ccc}{Forbes link} and CNET: \href{https://www.cnet.com/news/esa-spacex-starlink-satellite-nearly-collides-with-european-aeolus-satellit}{CNET link}

{[}58{]} Scientific American, "The FCC’s Approval of SpaceX’s Starlink Mega Constellation May Have Been Unlawful": \href{https://www.scientificamerican.com/article/the-fccs-approval-of-spacexs-starlink-mega-constellation-may-have-been-unlawful/}{Scientific American link}

{[}59{]} Microwave Journal, "5G Phased Array Technologies": \href{https://www.mathworks.com/content/dam/mathworks/ebook/microwave-journal-5g-phased-array-technologies-ebook.pdf}{phased-array-technologies-ebook}

{[}60{]} Anthony Mallama, "A  Flat-Panel Brightness Model for the Starlink Satellites and Measurement of their Absolute Visual Magnitude", \href{https://arxiv.org/abs/2003.07805}{https://arxiv.org/abs/2003.07805}

{[}61{]} A. Grazian, et al, "The LBC Exposure Time Calculator", \href{http://lbc.oa-roma.inaf.it/ETC/ETC_help/index.html}{lbc.oa-roma.inaf.it}

\end{document}